\documentclass[aps,pre,amsmath,amssymb,twocolumn,showpacs]{revtex4-1}
\usepackage{epsfig}
\usepackage{color}

\def\be{\begin{equation}}
\def\ee{\end{equation}}

\def\ba{{\bf a}}
\def\bb{{\bf b}}
\def\bE{{\bf e}}

\def\br{{\bf r}}
\def\bu{{\bf u}}

\def\bx{{\bf x}}

\def\bA{{\bf A}}
\def\bB{{\bf B}}

\def\bW{{\bf W}}
\def\wh{\widetilde}
\def\lb{\label}

\def\boLambda{\hbox{\boldmath $\Lambda$}}

\def\bdot{\hbox{\boldmath $\cdot$}}
\def\btimes{\hbox{\boldmath $\times$}}

\def\bzed{\hbox{\boldmath $0$}}
\def\grad{\hbox{\boldmath $\nabla$}}
\def\bell{\hbox{\boldmath $\ell$}}

\begin{document}


\title{Fluctuation Dynamo and Turbulent Induction at Small Prandtl Number}


\author{Gregory L. Eyink}
\email{eyink@jhu.edu}
\affiliation{%
Department of Applied Mathematics \& Statistics\\
and Department of Physics \& Astronomy\\
The Johns Hopkins University, USA}%


\date{\today}

\begin{abstract}
We study the Lagrangian mechanism of the fluctuation dynamo at zero Prandtl
number
and infinite magnetic Reynolds number, in the Kazantsev-Kraichnan model of
white-noise
advection. With a rough velocity field corresponding to a turbulent
inertial-range, flux-freezing
holds only in a stochastic sense. We show that field-lines arriving to the same
point which
were initially separated by many resistive lengths are important to the dynamo.
Magnetic
vectors of the seed field that point parallel to the initial separation vector
arrive anti-correlated
and produce an ``anti-dynamo'' effect. We also study the problem of  ``magnetic
induction''
of a spatially uniform seed field. We find no essential distinction between
this process and
fluctuation dynamo, both producing the same growth-rates and small-scale
magnetic
correlations. In the regime of very rough velocity fields where fluctuation
dynamo fails,
we obtain the induced magnetic energy spectra. We use these results to evaluate
theories
proposed for magnetic spectra in laboratory experiments of turbulent induction.
\end{abstract}


\maketitle

\section{Introduction}\lb{intro}

Hannes Alfv\'en introduced the notion of magnetic flux conservation in ideal
magnetohydrodynamics \cite{Alfven42}. This property of  ``flux-freezing''
involves an implicit
assumption, however, that plasma fluid velocities remain smooth in the limit of
vanishing
resistivity. This assumption need not be valid if the viscosity vanishes
together with
the resistivity, at a fixed or a decreasing magnetic Prandtl number. At the
implied
high Reynolds numbers, smooth laminar plasma flows will be unstable to  the
development of turbulence.  We have argued previously that magnetic
flux-conservation
in its usual sense cannot hold at any small resistivities,  no matter how tiny,
in a
turbulent plasma with a Kolmogorov-like energy spectrum \cite{EyinkAluie06}.
Neither is flux-freezing  completely broken but, instead,  it becomes
intrinsically
stochastic  due to the roughness of the advecting velocity field
\cite{Eyink07,Eyink09}.
Infinitely-many magnetic field lines are carried to each point---even in the
limit of very high
conductivity---which must be averaged to obtain the resultant magnetic field at
that point.

The previous ideas are theoretical predictions for turbulent flows of
astrophysical
and laboratory plasmas. However, they are exact properties of at least one
soluble problem of turbulent advection, the Kazantsev-Kraichnan (KK) model of
kinematic magnetic dynamo. See \cite{Kazantsev68,KraichnanNagarajan67,
Kraichnan68} and many following papers \cite{RuzmaikinSokolov81,
Novikovetal83,Vergassola96,RogachevskiiKleeorin97,
Vincenzi02,BoldyrevCattaneo04,Boldyrevetal05,Celanietal06,ArponenHorvai07}.
In that model the physical turbulent velocity field is replaced by a Gaussian
random
field that is white-noise in time and rough (H\"older continuous) in space. It
is rigorously
established for this model that Lagrangian particle trajectories are
``spontaneously
stochastic'' in the limit of infinite Reynolds numbers, with an infinite
ensemble of
trajectories for a given advecting velocity field and a given initial particle
position
\cite{Bernardetal98,GawedzkiVergassola00,Chavesetal03,EvandenEijnden00,
EvandenEijnden01,LeJanRaimond02,LeJanRaimond04,Kupiainen03}.
This remarkablebreakdown of Laplacian determinism is a consequence of the
properties of two-particle Richardson diffusion \cite{Richardson26}, which are
also
expected to hold for real fluid turbulence.

The stochastic nature of flux-freezing must play an essential role in the
turbulent
magnetic dynamo at high kinetic Reynolds numbers, either for fixed or for
vanishing
magnetic Prandtl numbers.  The zero Prandtl-number fluctuation dynamo was
studied
in an earlier work of  Eyink and Neto \cite{EyinkNeto10}, hereafter referred to
as I, primarily
for the KK model. This model undergoes a remarkable transition as the scaling
exponent
$\xi$ of the velocity structure function is decreased,  with dynamo effect
disappearing
for $\xi<1$ in space dimension three. This seems rather counterintuitive,
because
velocity-gradients and line-stretching rates in the model {\it increase} as
$\xi$ is lowered,
which should favor dynamo action. Paper I showed that the stochasticity of
flux-freezing is
fundamental to understand this transition. Although magnetic line-stretching
rates indeed
increase for decreasing $\xi,$ the resultant magnetic field at a point is the
average of
infinitely-many field lines advected to that point from a large spatial volume.
As shown
in I, the {\it correlations} between independent line-vectors are small for
$\xi<1,$ when
the velocity field is too rough, and the net magnetic field decays despite the
rapid growth
in strength of individual field lines.

The contribution of flux-line stochasticity to turbulent dynamo action for
$\xi>1$ was not fully
explicated in I, however.  Infinitely-many field lines are advected to each
point from a large
spatial volume, e.g. with diameter of order the velocity integral length $L$ in
one turbulent
turnover time $L/u_{rms}.$ But all of these field lines are unlikely to make an
equal
contribution to dynamo action. In particular, field vectors that arrive to the
same point
from very distant initial locations must be poorly correlated and give small
(or negative)
dynamo effect. It is not hard to guess that dynamo action in the limit of
high-Reynolds number
$Lu_{rms}/\nu \gg 1$ and low Prandtl number $\nu/\eta\ll 1$ must arise mainly
from vectors with
initial separation distances $r$ of order $\sim \ell_\eta,$ the resistive
length. Indeed, this is the only
available length-scale in that limit, on dimensional grounds. The exact
contribution to fluctuation
dynamo from vectors with initial separation $r$ remains unknown, however, even
in the KK model.
One of the main purposes of this paper to provide a quantitative answer to this
problem.

Another important issue left unresolved in I was the outcome of a
spatially-uniform
initial magnetic field in the dynamo regime of the KK model for $\xi>1.$
Can such a magnetic field provide the seed for a small-scale
fluctuation-dynamo? This
question is very closely connected with the proper formulation of  the concept
of
``magnetic induction'' . This process is usually defined as the creation of
small-scale magnetic
fluctuations by the turbulent ``shredding'' of a non-zero mean magnetic field
and has been
invoked to explain small-scale magnetic fluctuations in liquid metal
experiments
\cite{Odieretal98,Bourgoinetal02,Peffleyetal00,Spenceetal06,Nornbergetal06},
in related numerical simulations \cite{Baylissetal07,Schekochihinetal07}, and
at the solar surface \cite{BrandenburgSubramanian05,SchuesslerVoegler08}.
It is frequently suggested that this process is distinct from fluctuation
dynamo. For example,
Schekochihin et al. wrote: ``Given a multiscale observed or simulated magnetic
field, one
does not generally have enough information (or understanding), to tell whether
it has originated
from the fluctuation dynamo, from the mean-field dynamo plus the turbulent
induction or from
some combination of the two.'' \cite{Schekochihinetal07}  However, if a mean
magnetic
field can provide a seed for fluctuation dynamo, is there really any meaningful
physical
distinction between the two mechanisms? Or if magnetic induction is defined
instead
as a process that occurs only in the absence of fluctuation dynamo, as it
sometimes is,
then how can one regard the two processes as acting in ``combination"? Clearly
there
is some conceptual confusion in the literature about the proper definition of
magnetic
induction. We shall use our exact results in the KK model to discuss these
issues
and, also, to evaluate some the proposed theories of the induced magnetic
spectrum.

The contents of this paper are as follows. In section \ref{back} we provide
more detailed background to our work and a more formal statement of the
problems
to be studied. In section \ref{dynamo} we study the Lagrangian mechanism of the
fluctuation dynamo at zero Prandtl number and the quantitative role of
stochastic flux-freezing.
A first subsection \ref{analysis} presents exact mathematical analysis and the
second subsection
\ref{numerical} gives numerical results. The next section \ref{induction}
discusses
the problem of magnetic induction using our analytical results for the KK
model. We draw
our final conclusions in section \ref{conclusions}. Appendices \ref{BCR} and
\ref{numerics} present some technical material.

\section{Background and Problem Statement}\lb{back}

We begin this section with a summary of the key results of Paper I.
 A ``dynamo order-parameter''  was introduced there with purely geometric
significance. Let $\bx(\ba,t)$ be a stochastic Lagrangian trajectory,
satisfying
\be d\bx = \bu(\bx,t)dt + \sqrt{2\eta}\,d\bW(t),\,\,\,\,\,\,\bx(t_0)=\ba.
\lb{x-def} \ee
We shall assume in this section that the advecting velocity field is
incompressible,
$\grad\bdot\bu=0,$ as in I.  $\bW(t)$ in (\ref{x-def}) is a vector  Brownian
motion
and $\eta$ is the resistivity (or, in fact, the magnetic diffusivity). Let
$\bell_k(\ba,t)$
be a passive vector starting as a unit vector $\hat{\bE}_k$ at space point
$\ba$ and time
$t_0$ and subsequently transported along $\bx(\ba,t),$ stretched and rotated by
the velocity-gradient. That is,
\be \bell_k(\ba,t) = \hat{\bE}_k\bdot\grad_a\bx(\ba,t). \lb{bell-def} \ee
Then I introduced the quantity
\be  \mathcal{R}_{k\ell}(\br,t)
=\langle\overline{\bell_k(\ba,t)\bdot\bell_\ell'(\ba',t)
     \delta^3(\bx(\ba,t)-\bx'(\ba',t))}\rangle, \lb{R-def} \ee
with $\br=\ba-\ba',$ where $\langle\cdot\rangle$ denotes average over the
turbulent velocity
realizations, $\overline{(\cdot)}$ denotes average over the random Brownian
motions and $\prime$ indicates a second, independent realization of the latter.
This quantity measures not only the magnitude but also the angular correlation
of independent line-vectors that start as unit vectors
$\bell_k(\ba,t_0)=\bE_k,\,
\bell_\ell'(\ba',t_0)=\bE_\ell$ at points $\ba,\ba'$ separated by displacement
$\br,$
which end at the {\it same} point at time $t.$

Now let $\bB$ be a magnetic field kinematically transported by the random
velocity field
$\bu$ and diffused by
resistivity $\eta$:
\be \partial_t\bB+(\bu\bdot\grad)\bB=(\bB\bdot\grad)\bu +\eta\triangle\bB,
 \,\,\,\,\,\,\bB(t_0)=\bB_{(0)}.  \lb{B-eq1} \ee
As shown in I, the mean energy in this magnetic field can be expressed as
\be \langle B^2(t)\rangle = \int  d^3r \, \mathcal{R}_{k\ell}(\br,t)
       \langle B^k_{(0)}(\br) B^\ell_{(0)}(\bzed)\rangle.  \lb{Ben-R-rel} \ee
It has been assumed in deriving (\ref{Ben-R-rel}) that initial conditions on
the magnetic
field are statistically independent of the velocity and that both quantities
are
statistically homogeneous (space translation-invariant). This formula makes
clear the close relation of dynamo action to the line-correlation
(\ref{R-def}), which grows exponentially rapidly precisely in the dynamo
regime.

This connection was verified in I by an exact calculation of  the
line-correlation
(\ref{R-def}) in the soluble Kazantsev-Kraichnan (KK) model of turbulent dynamo
\cite{Kazantsev68,KraichnanNagarajan67,Kraichnan68,Vergassola96,
RogachevskiiKleeorin97,Vincenzi02,BoldyrevCattaneo04,
Boldyrevetal05,Celanietal06,ArponenHorvai07}.  This model replaces
the true turbulent velocity by  a Gaussian random field which
is white-noise in time:
\be \langle u^i(\bx,t) u^j(\bx',t')\rangle=\kappa^{ij}(\br)\delta(t-t')
\lb{u-corr} \ee
with $\br=\bx-\bx'.$
In this model, the magnetic correlation function
\be \mathcal{C}^{ij}(\br,t)=\langle B^i(\br,t) B^j(\bzed,t)\rangle \lb{C-def}
\ee
evolves according to a closed, linear equation
\be \partial_t\mathcal{C}^{ij}= \mathcal{M}^{ij}_{k\ell}\mathcal{C}^{k\ell},
\lb{C-eq} \ee
where $\mathcal{M}$ is a singular diffusion operator. If $\kappa^{ij}(\br)\sim
r^\xi$ for small $r,$ with rugosity exponent $0<\xi<2,$ then the velocity
realizations are rough in space and provide a qualitatively good model of
turbulent advection. It was shown in I
that $\mathcal{R}_{k\ell}(\br,t)$ in the KK model is the solution of the
adjoint
equation
\be \partial_t\mathcal{R}_{k\ell}= \mathcal{M}_{k\ell}^{ij\,*}\mathcal{R}_{ij}
\lb{R-eq} \ee
with initial condition $\mathcal{R}_{k\ell}(\br,t)=\delta_{k\ell}
\delta^3(\br).$
This fact was exploited to show that $\mathcal{R}_{k\ell}(\br,t)$ in the KK
model
at zero magnetic Prandtl number, as expected, grows exponentially in the
dynamo regime ($\xi>1$) but has only power-law time-dependence in the
non-dynamo regime ($\xi<1$).

The study in I left open, however, several important questions. First,
what range of $r=|\br|$ substantially contributes to the correlation
$\mathcal{R}$ and to the space-integral in (\ref{Ben-R-rel})? It is known
in the KK model that magnetic field lines arrive to a point from a spatial
region with radius $L(t)\sim (t-t_0)^{1/(2-\xi)},$ in the root-mean-square
sense. This is also expected for kinematic dynamo in a real turbulent
fluid, where $\xi\doteq 4/3$ and $L(t)\sim (t-t_0)^{3/2}$ corresponds
to turbulent Richardson diffusion of the field lines. However, not all
of the field lines in this large volume should be expected to contribute
equally to dynamo action. Field vectors starting from points separated
by distances $r$ much larger than the resistive length-scale may
arrive at the same point with small angular correlations and thus
contribute little to net magnetic field growth. It follows from the analysis
of I for the KK model in its dynamo regime that, for times $t$ long compared
to the resistive time-scale,
\be \mathcal{R}_{k\ell}(\br,t) \sim e^{-\lambda t}
\mathcal{\wh{G}}_{k\ell}(\br),
\,\,\,\,\,\,\,\,\,\, (\lambda<0) \lb{R-longT} \ee
where $-\lambda$ is the largest (positive) eigenvalue of $\mathcal{M}^*$ and
$\mathcal{\wh{G}}$ is the corresponding (right) eigenmode. Thus, questions
about the spatial structure of $\mathcal{R}$ reduce, in the exponential
growth regime at sufficiently long times, to the corresponding questions
about the right eigenmode $\mathcal{\wh{G}}$ of $\mathcal{M}^*$
(or, equivalently, left eigenmode of $\mathcal{M}$). It is one of the goals
of the present work to calculate this eigenmode and thereby determine
the degree of correlation of line-vectors arriving from various initial
separations.

A second important issue left unresolved in I was the response of a
spatially uniform initial magnetic field to turbulent advection and stretching.
In the case of a uniform initial field, formula (\ref{Ben-R-rel}) simplifies to
\be \langle B^2(t)\rangle =  \mathcal{R}_{k\ell}(t)\langle
B^k_{(0)}B^\ell_{(0)}\rangle,
\lb{Ben-R-hom} \ee
with
\be \mathcal{R}_{k\ell}(t)\equiv \int d^3r\,\mathcal{R}_{k\ell}(\br,t)
    = \langle\bell_k(t)\bdot\bell_\ell'(t)\rangle_0. \lb{Rint-def} \ee
The quantity (\ref{Rint-def}) measures the correlation of two independent
line-vectors
that arrive at the same point, regardless of their initial separation. The
result
(\ref{R-longT}) does not imply exponential growth of this integrated
correlation,
however, unless one can show that
\be  \int d^3r \, \mathcal{\wh{G}}_{k\ell}(\br)\neq 0. \ee
Below we shall demonstrate this fact.

This result is closely connected with another important physical problem, the
creation of small-scale magnetic fluctuations by turbulent ``magnetic
induction''
of a non-zero mean-field \cite{Odieretal98,Bourgoinetal02,Peffleyetal00,
Spenceetal06,Nornbergetal06,Baylissetal07,Schekochihinetal07}. Indeed,
the formula (\ref{Ben-R-hom}) applies directly to this situation, which
corresponds to the special case $\langle B^k_{(0)} B^\ell_{(0)}\rangle
=\langle B^k_{(0)}\rangle \langle B^\ell_{(0)}\rangle.$  This observation
already makes clear that there is no essential physical distinction
between ``fluctuation dynamo'' and ``magnetic induction'', which simply
correspond
to different choices of initial magnetic field (random vs. deterministic).
Other definitions
of ``magnetic induction'' are offered in the literature. If the fluctuation
dynamo
fails to operate for some reason (e.g. if the magnetic Reynolds number is too
small),
then the term ``magnetic induction'' is sometimes used for the process by which
small-scale fluctuations are generated from the mean magnetic field. If a
mean-field
dynamo still operates and creates an exponentially growing large-scale magnetic
field,
then these ``parasitic'' small-scale fluctuations may be also exponentially
increasing.
This provides another mechanism to produce exponential growth of small-scale
magnetic fields in the absence of a fluctuation dynamo. If instead  there is no
mean-field dynamo, then such growth of small-scale fluctuations by ``magnetic
induction''
is sub-exponential.  We shall discuss below the non-dynamo regime $(\xi<1)$ of
the
KK model as a simple example of  such ``magnetic induction'' and use it to
evaluate
some physical theories that have been proposed for the phenomenon.

We mention finally one additional motivation to study the left eigenmodes
of $\mathcal{M}.$ As pointed out in I, these
have a physical interpretation as correlators of the magnetic vector potential
$\bA.$
More precisely, the correlation function defined by
\be \mathcal{G}_{k\ell}(\br,t)=\langle A_k(\br,t) A_\ell(\bzed,t)\rangle,
\lb{G-def} \ee
in the KK model satisfies
\be \partial_t\mathcal{G}_{k\ell}= \mathcal{M}_{k\ell}^{ij\,*}\mathcal{G}_{ij},
\lb{G-eq} \ee
with diffusion operator adjoint to that in eq.(\ref{C-eq}). The underlying
reason for this
fact is the conservation of magnetic helicity $H=\int d^3r\, \bA\bdot\bB$ for
the ideal
induction equation. If the latter is written as
\be \partial_t \bB = \mathcal{L}\bB, \lb{B-Leq} \ee
for a linear operator $\mathcal{L}$ (depending on the velocity $\bu$), then
helicity
conservation is equivalent to
\be \partial_t\bA = -\mathcal{L}^*\bA,  \lb{A-Leq} \ee
up to addition of a gradient term. Different operators $\mathcal{L}$ are
possible, which coincide in their action
on solenoidal magnetic fields ($\grad\bdot\bB=0$), but whose adjoints
$\mathcal{L}^*$
correspond to different gauge choices for $\bA.$  Equation (\ref{A-Leq})
directly
implies (\ref{G-eq}) in the KK model. The addition of resistivity to eqs.
(\ref{B-Leq}),
(\ref{A-Leq}) does not change this result, since the additional Laplacian
operator is
self-adjoint.

\section{Fluctuation Dynamo}\lb{dynamo}

We study in this section the Lagrangian mechanism of the fluctuation dynamo in
the KK model for three space-dimensions (3D) and for homogeneous and isotropic
statistics of both velocity and magnetic fields. We begin in the first
subsection \ref{analysis}
with exact mathematical analysis of the problem.  Most of this discussion
applies to a fairly
general situation, allowing for compressibility and helicity of the velocity
field. We then
specialize to the case of  an incompressible, non-helical advecting velocity,
with
a power-law space-correlation corresponding to an infinite-Reynolds-number
inertial-range range for the velocity field. In the second subsection
\ref{numerical} we
present numerical results for this latter case and discuss their physical
interpretation.
The reader who is mostly interested in the final results, and not their
detailed
derivation, may skip directly to that section of the paper.

\subsection{Mathematical Analysis}\lb{analysis}

We discuss two natural choices for the linear operator $\mathcal{L}$ in
the ideal induction eq.(\ref{B-Leq}) and the corresponding adjoint operators
$\mathcal{L}^*$ and gauge choices for the vector potential, successively
in the following two subsections.

\subsubsection{Gauge I}\lb{analytic-I}

One choice to write the induction equation is as
\be \partial_t\bB=\grad\btimes(\bu\btimes\bB-\eta\grad\btimes\bB). \lb{B-eq-I}
\ee
Note that in the ideal equation with $\eta=0,$ $\mathcal{L}_\bu^{(2)}\bB=
-\grad\btimes(\bu\btimes\bB)$ is the Lie-derivative acting on $\bB$ as a
differential 2-form. See \cite{Larsson03}, eq.(2.3) for $\grad\bdot\bB=0.$
Thus, this form of the induction equation is most directly connected with the
conservation of magnetic flux through advected 2-dimensional surfaces.
The corresponding adjoint equation for the vector potential is
\begin{eqnarray}
\partial_t\bA &=& \bu\btimes(\grad\btimes\bA)-\eta\grad\btimes(\grad\btimes\bA)
\cr
                       &=&-(\bu\bdot\grad)\bA + (\grad\bA)\bu \cr
&&\,\,\,\,\,\,\,\,\,\,\,\,\,\,\,\,\,\,\,\,\,\,\,\,\,\,\,\,\,\,\,
                        +\eta[\triangle\bA-\grad(\grad\bdot\bA)].
\lb{A-eq-I} \end{eqnarray}
In the gauge-choice implied by this equation,  pure-gauge fields
$\bA=\grad\lambda$
satisfy $\grad(\partial_t\lambda)=0$ and are, thus, time-independent up to
possible
spatially constant terms.

Boldyrev, Cattaneo and Rosner in \cite{Boldyrevetal05} developed an elegant
formulation of the KK model based on the eq.(\ref{B-eq-I}) for the case of
homogeneous
and isotropic statistics. We review their formulation---hereafter referred to
as the
BCR formalism---in Appendix \ref{BCR}, paying special attention to
issues of gauge-invariance. Readers not previously familiar with the work of
\cite{Boldyrevetal05} may wish to review that appendix before proceeding. One
essential observation of BCR is that the evolution operator for magnetic
correlations
{\it factorizes} as ${\cal M}={\cal DJ}.$ More precisely, in the KK model
starting from
eq.(\ref{B-eq-I})
\be \partial_t \mathcal{C}^{ij}= \mathcal{M}^{ij}_{pq}\mathcal{C}^{pq}
=\mathcal{D}^{ij,k\ell}\mathcal{J}_{k\ell,pq}\mathcal{C}^{pq} \lb{C-eq-long}
\ee
where
\be \mathcal{ D}^{ij,k\ell}=\epsilon^{ikp}\epsilon^{j\ell
q}\partial_p\partial_q
\lb{D-def} \ee
is the non-positive, self-adjoint differential operator which relates the
magnetic
correlation to the vector-potential correlation and
\be \mathcal{J}_{ij,k\ell}=\epsilon_{ikp}\epsilon_{j\ell q}T^{pq} \lb{J-def}
\ee
is the self-adjoint multiplication operator with
$ T^{pq}(\br)= 2\eta\delta^{pq} + \kappa^{pq}(0)-\kappa^{pq}(\br), $
where $\kappa^{pq}(\br)$ is the spatial velocity-correlation in
eq.(\ref{u-corr}).
Explicitly,
\begin{eqnarray}
\mathcal{M}^{ij}_{pq}\mathcal{C}^{pq} &=&
\partial_r\partial_s\left(T^{rs}\mathcal{C}^{ij}\right)
-\partial_p\partial_s\left(T^{is}\mathcal{C}^{pj}\right) \cr
&& \,\,\,\,\,-\partial_r\partial_q\left(T^{rj}\mathcal{C}^{iq}\right)
+\partial_p\partial_q\left(T^{ij}\mathcal{C}^{pq}\right).
\lb{M-def-I} \end{eqnarray}

One may further write equations in the KK model for the joint correlations of
magnetic and vector-potentials
$$ \Psi^{i\,\,\,\,}_{\,\,\,\,k}(\br,t) = \langle B^i(\br,t)A_k(\bzed,t)\rangle,
\,\,\Psi_{k\,\,\,\,}^{\,\,\,\,i}(\br,t) = \langle
A_k(\br,t)B^i(\bzed,t)\rangle$$
using the operator $\mathcal{R}^{i,k} = \epsilon^{i p k}\partial_p,$
in terms of which the operator $\mathcal{D}$ itself factorizes as $\mathcal{
D}^{ij,k\ell}
=\mathcal{R}^{i,k}\mathcal{R}^{j,\ell}.$ Then
\begin{eqnarray}
 \partial_t \Psi^{i\,\,\,\,}_{\,\,\,\,n} &=&
-(\mathcal{R}^{i,m})^*\mathcal{J}_{mn,pq}
     \mathcal{R}^{q,\ell}\Psi^{p\,\,\,\,}_{\,\,\,\,\ell} \cr
\partial_t \Psi_{m\,\,\,\,}^{\,\,\,\,j} &=&
-(\mathcal{R}^{j,n})^*\mathcal{J}_{mn,pq}
     \mathcal{R}^{p,k}\Psi_{k\,\,\,\,}^{\,\,\,\,q} \lb{Psi-eq}
\end{eqnarray}
Finally, the vector-potential correlation in (\ref{G-def}) obeys
\be \partial_t \mathcal{G}_{k\ell}=
(\mathcal{M}^*)_{k\ell}^{pq}\mathcal{G}_{pq}
=\mathcal{J}_{k\ell,rs}\mathcal{D}^{rs,pq}\mathcal{G}_{pq} \lb{G-eq-I} \ee
with $\mathcal{M}^*=\mathcal{JD}.$ Explicitly,
\begin{eqnarray}
(\mathcal{M}^*)_{k\ell}^{pq}\mathcal{G}_{pq} &=&
T^{rs}\partial_r\partial_s\mathcal{G}_{k\ell}
-T^{ps}\partial_k\partial_s\mathcal{G}_{p\ell} \cr
&& \,\,\,\,\,-T^{rq}\partial_r\partial_\ell\mathcal{G}_{kq}
+T^{pq}\partial_k\partial_\ell\mathcal{G}_{pq}.
\lb{Mstar-def-I} \end{eqnarray}

BCR mainly considered the isotropic sector of the KK model, invariant
under proper rotations but not necessarily space reflections. In that case,
with $\hat{\br}=\br/r,$
$$ \kappa^{ij}(\br)=\kappa_L(r) \hat{r}^i\hat{r}^j
       +\kappa_N(r)(\delta^{ij}-\hat{r}^i\hat{r}^j)+ \kappa_H(r) \epsilon^{ijk}
\hat{r}_k, $$
where $\kappa_L,\,\,\kappa_N,$ and $\kappa_H=r g$ are conventional
longitudinal, transverse and helical correlation functions, with similar
representations
of $\mathcal{C},\mathcal{G},$ etc. However, ref.\cite{Boldyrevetal05}
introduced instead a special basis to expand tensor correlation functions, as
$\mathcal{C}^{ij}=\sum_{a=1}^3 C_a \xi^{ij}_a$ with
$$ \xi^{ij}_1 = {{1}\over{\sqrt{2}r}}(\delta^{ij}-\hat{r}^i\hat{r}^j),\,\,\,\,
      \xi^{ij}_2 = {{1}\over{r}}\hat{r}^i\hat{r}^j,\,\,\,\,
      \xi^{ij}_3 = {{1}\over{\sqrt{2}r}}\epsilon^{ijk}\hat{r}^k, $$
so that
$$ C_1=\sqrt{2}rC_N,\,\,\,\,C_2=rC_L,\,\,\,\,C_3=\sqrt{2}r C_H. $$
The special feature of this basis is that the Hilbert-space
inner-product defined by
$$ \langle \mathcal{G,C}\rangle = \int d^3 r\,\,
\mathcal{G}_{ij}(\br) \mathcal{C}^{ij}(\br) $$
simplifies in the isotropic sector to
$$
\langle G,C\rangle =4\pi \int_0^\infty dr\,[G_1C_1+G_2C_2+G_3C_3]. $$
We hereafter use the notation $\mathcal{C,G}$ etc. in the BCR formalism for the
isotropic
sector to represent the column vectors
$$      \mathcal{C}=\left(\begin{array}{c}
                             C_1\cr
                             C_2\cr
                             C_3
                             \end{array}\right). $$
In this representation, the operators ${\cal J}$ and ${\cal D}$ take the simple
forms
$$ \mathcal{J}=\left(\begin{array}{ccc}
                             b & a & 0\cr
                             a & 0 & c \cr
                             0 & c & b
                             \end{array}\right) ,\,\,\,\,\,
      \mathcal{D}=\left(\begin{array}{ccc}
                             \partial_r^2 & -\partial_r{{\sqrt{2}}\over{r}} &
0\cr
                             {{\sqrt{2}}\over{r}}\partial_r & -{{2}\over{r^2}}
& 0 \cr
                0 & 0 & {{1}\over{r^2}}\partial_r  r^4\partial_r{{1}\over{r^2}}
                             \end{array}\right), $$
with
$$ a(r)=\sqrt{2}[2\eta+\kappa_N(0)-\kappa_N(r)]$$
$$ b(r)=2\eta+\kappa_L(0)-\kappa_L(r)$$
$$ c(r)=\sqrt{2}[g(0)-g(r)]r. $$

A crucial observation of BCR is that $\mathcal{D}$ factorizes as
$\mathcal{D}=-\mathcal{RR}^*$ with
$$       \mathcal{R}=\left(\begin{array}{ccc}
                                            0 & \partial_r &  0\cr
                                            0 & {{\sqrt{2}}\over{r}} & 0 \cr
                                            0 & 0 & -{{1}\over{r^2}}\partial_r
r^2
                                            \end{array}\right), \,\,\,\,\,\,
        \mathcal{R}^*=\left(\begin{array}{ccc}
                                             0 & 0  &  0\cr
                                            -\partial_r & {{\sqrt{2}}\over{r}}
& 0 \cr
                                            0 & 0 &
r^2\partial_r{{1}\over{r^2}}
                                            \end{array}\right).  $$
Another important fact is that ${\rm Ran}(\mathcal{R}),$ the range of the
operator $\mathcal{R},$
consists of the solenoidal functions $\mathcal{C}^{ij}$ that satisfy
$\partial_i\mathcal{C}^{ij}=
\partial_j \mathcal{C}^{ij}=0.$  Thus, all solenoidal solutions of the equation
$\partial_t
\mathcal{C}=\mathcal{MC}=-\mathcal{RR}^*\mathcal{JC}$ can be obtained by
solving
\be \partial_t W= - \mathcal{R}^*\mathcal{JR} W \lb{W-eq} \ee
with a self-adjoint operator $\mathcal{S}=\mathcal{R}^*\mathcal{JR}$, and then
setting
$$ \mathcal{C}=\mathcal{R}W=\left(\begin{array}{c}
                                             \partial_rW_2 \cr
                                             {{\sqrt{2}}\over{r}}W_2 \cr
                                            -{{1}\over{r^2}}\partial_r( r^2W_3)
                                            \end{array}\right). $$
Comparison of eq.(\ref{W-eq}) in the isotropic sector with the general
eq.(\ref{Psi-eq})
suggests that $W$ is a simple linear transformation of the magnetic-field,
vector-potential correlation
$\Psi.$ For this result, and for the derivation of all the preceding
statements,
see Appendix \ref{BCR}.

We now discuss the solutions of the adjoint problem $\partial_t
\mathcal{G}=\mathcal{M}^*
\mathcal{G}=-\mathcal{JRR}^*\mathcal{G}.$  Its solutions are likewise related
to solutions
of (\ref{W-eq}) by the equation
$$W=\mathcal{R}^*\mathcal{G}=\left(\begin{array}{c}
                                             0\cr
                                            -\partial_rG_1
+{{\sqrt{2}}\over{r}}G_2\cr

r^2\partial_r\left({{G_3}\over{r^2}}\right)
                                            \end{array}\right). $$
This relation is many-to-one. Since ${\rm Ker}(\mathcal{R}^*)=
\left[{\rm Ran}(\mathcal{R})\right]^\perp,$
the kernel of $\mathcal{R}^*$ consists of the correlations of gradient type:
\be G_1(r) =\sqrt{2}\Lambda'(r),\,\,\,\,G_2(r) = r\Lambda''(r), \lb{gradient}
\ee
for a scalar correlation function $\Lambda.$ Thus, any solutions $\mathcal{G}$
of the adjoint
equation that differ by such a gradient solution are mapped to the same
solution $W$ of (\ref{W-eq}).
This freedom just corresponds to gauge-invariance of $W.$

We next employ this formalism to discuss the right eigenfunctions
$\mathcal{C}_\alpha$ and
left eigenfunctions $\mathcal{G}_\alpha$ of $\mathcal{M},$ which satisfy
$$ \mathcal{MC}_\alpha=-\lambda_\alpha\mathcal{C}_\alpha,\,\,\,\,\,
     \mathcal{M}^*\mathcal{G}_\alpha=-\lambda_\alpha\mathcal{G}_\alpha,$$
respectively. (Of course, for the continuous spectrum of the operators these
are generalized
eigenfunctions.) Let us suppose that one has obtained the eigenfunction
$W_\alpha$
of the self-adjoint operator $\mathcal{S}=\mathcal{R}^*\mathcal{JR}$:
\be \mathcal{S}W_\alpha = \mathcal{R}^*\mathcal{JR}W_\alpha =
                         \lambda_\alpha W_\alpha. \lb{S-eig} \ee
Then it is not hard to see that
\be \mathcal{C}_\alpha =\mathcal{R}W_\alpha,\,\,\,\,
     \mathcal{G}_\alpha={{1}\over{\lambda_\alpha}}\mathcal{JR}W_\alpha.
     \lb{eig-funs} \ee
The first result is obvious. To obtain the second, note that, with
$\mathcal{G}_\alpha$ as defined above,
$$ \mathcal{R}^* \mathcal{G}_\alpha={{1}\over{\lambda_\alpha}}
     \mathcal{S}W_\alpha=W_\alpha.$$
If this result is used to eliminate $W_\alpha$ in the definition of
$\mathcal{G}_\alpha,$ then
the statement follows. An interesting consequence is that right and left
eigenfunctions
are related very simply by
$$ \mathcal{G}_\alpha={{1}\over{\lambda_\alpha}}\mathcal{J}\mathcal{C}_\alpha.
$$
Note also that $\langle\mathcal{C}_\beta,\mathcal{G}_\alpha\rangle
=\langle \mathcal{R}W_\beta,
{{1}\over{\lambda_\alpha}}\mathcal{JR}W_\alpha\rangle
={{1}\over{\lambda_\alpha}}\langle W_\beta, \mathcal{S}W_\alpha\rangle=\langle
W_\beta,W_\alpha\rangle=\delta_{\alpha,\beta}.$ Thus, the left and right
eigenfunctions
are biorthogonal.

Our discussion so far has been fairly general, permitting a compressible
velocity field
with reflection-non-symmetric (helical) statistics. However, we now specialize.
If
space-reflection symmetric statistics are assumed, then $W_3=C_3=G_3=0.$
Furthermore,
$c(r)=0$ in the operator ${\cal J}.$ Thus, the eigenvalue problem (\ref{S-eig})
reduces to a single
equation for $W_2$:
$$ -\partial_r(b(r)\partial_rW_2) +
{{\sqrt{2}}\over{r^2}}[a(r)-ra'(r)]W_2=\lambda W_2. $$
This is a standard Sturm-Liouville eigenvalue problem. The relations
(\ref{eig-funs}) yield
\be C_1=\partial_r W_2,\,\,\,\, C_2 =  {{\sqrt{2}}\over{r}} W_2 \lb{R-eigfun}
\ee
(true even in the reflection non-symmetric case) and
\be G_1={{\sqrt{2}a(r)}\over{\lambda r}} W_2 + {{b(r)}\over{\lambda}}\partial_r
W_2,
      \,\,\,\,  G_2 =  {{a(r)}\over{\lambda}}\partial_r W_2. \lb{L-eigfun} \ee
The problem further simplifies if one assumes an incompressible flow, implying
the
relation $a=\sqrt{2}\left(b+{{1}\over{2}}rb'\right).$ In that case the
Sturm-Liouville problem
becomes
\be -\partial_r(p(r)\partial_rW_2) + q(r) W_2=\lambda W_2. \lb{SL-eq} \ee
with $p(r)=b(r)$ and
$$ q(r)= {{2b(r)-2rb'(r)-r^2b''(r)}\over{r^2}}=
      -\partial_r\left[{{1}\over{r^2}}\partial_r\left(r^2 b(r)\right)\right].
$$
{}From general Sturm-Liouville theory, the spectrum is bounded below and may
consist of both point spectrum and continuous spectrum, depending upon the
choice of the
function $b(r).$ Dynamo effect corresponds to the lowest  eigenvalue being
negative,  $\lambda<0,$
with a ground state, square-integrable wavefunction $\int_0^\infty
dr\,W_2^2(r)<\infty.$  The ground-state
eigenfunction $W_2$, if it exists at all, must be positive for all $r>0$ by
the Sturm oscillation theorem.

\subsubsection{Gauge II} \lb{analytic-II}

There is another natural choice for the linear operator $\mathcal{L}$ in
the ideal induction eq.(\ref{B-Leq}). This choice corresponds to writing that
equation as
\be \partial_t \bB = -(\bu\bdot\grad)\bB+(\bB\bdot\grad)\bu -(\grad\bdot\bu)\bB
+\eta\triangle\bB. \lb{B-eq-II} \ee
If $\bB$ is a smooth solution of eq. (\ref{B-eq-II}) for $\eta=0$ and if $\rho$
is the
mass density solving the continuity equation,
$\partial_t\rho+\grad\bdot(\rho\bu)=0,$
then, as is well-known, $\bell=\bB/\rho$ is a ``frozen-in'' field. That is,
$$ \frac{\partial}{\partial t}\bell=-(\bu\bdot\grad)\bell+(\bell\bdot\grad)\bu.
$$
The righthand side of this equation is just $\mathcal{L}^\bu_{(1)}\bell,$ the
Lie-derivative
operator acting upon vectors (rank-1 contravariant tensors) \cite{Larsson03}.
These
facts explain the interest of the eq. (\ref{B-eq-II}), since intuitive
geometric notions of
material line-vector motion can be exploited to understand magnetic dynamo
effect.
The results in I all depended upon using this form of the induction equation.

The equation for the vector-potential adjoint to (\ref{B-eq-II}) is
\be \partial_t \wh{\bA} =-(\bu\bdot\grad)\wh{\bA} -(\grad\bu)\wh{\bA}
+\eta\triangle\wh{\bA}.
\lb{A-eq-II} \ee
We use the tilde to distinguish the vector potential in this gauge from that
resulting
from (\ref{A-eq-I}). There is also a geometric significance to
eq.(\ref{A-eq-II}), since
$\mathcal{L}_\bu^{(1)}\wh{\bA}=(\bu\bdot\grad)\wh{\bA}+(\grad\bu)\wh{\bA}$ is
the Lie-derivative acting on $\wh{\bA}$ as a differential 1-form. This is
related
to the fact that ``frozen-in'' magnetic flux through surfaces (2-cells) can be
written
as the line-integral of the vector potential $\wh{\bA}$ (or $\bA$) around
closed loops
(1-cycles). In the gauge choice corresponding to (\ref{A-eq-II}) the pure-gauge
fields
$\wh{\bA}=\grad\wh{\lambda}$ have zero Lagrangian time-derivative,
$D_t\wh{\lambda}
=0,$ up to possible spatial constants. Of course, the two gauge choices $\bA$
and
$\wh{\bA}$ must be related by a suitable gauge transformation
$$ \wh{\bA}=\bA - \grad\lambda. $$
In fact, eq.(\ref{A-eq-II}) can be rewritten as
$$ \partial_t \wh{\bA} = \bu\btimes(\grad\btimes\wh{\bA})
     -\eta\grad\btimes(\grad\btimes\wh{\bA}) -\grad\Phi, $$
with
$$ \Phi=\bu\bdot\wh{\bA}-\eta\grad\bdot\wh{\bA}, $$
implying that $\lambda=\int^t dt'\,\,\Phi(t').$

In the KK model, the form of the induction equation (\ref{B-eq-II}) leads to
the
evolution equation for the magnetic correlation $\partial_t\mathcal{C}^{ij}=
\mathcal{\wh{M}}^{ij}_{pq}\mathcal{C}^{pq},$ with
\begin{eqnarray}
\mathcal{\wh{M}}^{ij}_{pq}\mathcal{C}^{pq} &=&
\partial_r\partial_s\left(T^{rs}\mathcal{C}^{ij}\right)
-\partial_s\left(\partial_pT^{is}\,\mathcal{C}^{pj}\right)\cr
& & \,\,-\partial_r\left(\partial_q T^{rj}\,\mathcal{C}^{iq}\right)
+   (\partial_p\partial_q T^{ij})\mathcal{C}^{pq}.
\lb{M-def-II} \end{eqnarray}
Comparison with the definition of $\mathcal{M}$ in (\ref{M-def-I}) shows that
$\mathcal{MC}=\mathcal{\wh{M}C}$ when acting on elements $\mathcal{C}$
of the subspace of solenoidal correlations functions. We note furthermore
that this subspace is invariant under the evolution (\ref{C-eq}) (because
the induction equation (\ref{B-eq-I}) preserves solenoidality of $\bB.$)
The adjoint equation for the vector-potential correlation
$\mathcal{\wh{G}}_{ij}(\br,t)=\langle\wh{A}_i(\br,t)\wh{A}_j(\bzed,t)\rangle$
is
\be \partial_t\wh{\mathcal{G}}_{k\ell}= (\wh{\mathcal{M}}^*)_{k\ell}^{pq}
\wh{\mathcal{G}}_{pq}, \lb{G-eq-II} \ee
with
\begin{eqnarray}
(\mathcal{\wh{M}}^*)_{k\ell}^{pq}\wh{\mathcal{G}}_{pq} &=&
T^{rs}\partial_r\partial_s\wh{\mathcal{G}}_{k\ell}
+ (\partial_k T^{ps})\partial_s\wh{\mathcal{G}}_{p\ell}\cr
& & \,\,+(\partial_\ell T^{rq})\partial_r\wh{\mathcal{G}}_{kq}
+(\partial_k\partial_\ell T^{pq})\wh{\mathcal{G}}_{pq}.
\lb{Mstar-def-II} \end{eqnarray}

In the previous section we have discussed how to determine the
eigenvalues and eigenfunctions of the operator $\mathcal{M}.$ The right
eigenfunctions $\mathcal{\wh{C}}_\alpha$ of $\mathcal{\wh{M}}$ and
$\mathcal{C}_\alpha$ of $\mathcal{M}$ are the same in the solenoidal
subspace, since $\mathcal{\wh{M}}=\mathcal{M}$ there.  We shall obtain
the left eigenfunctions of $\mathcal{\wh{M}}$ by solving for the differences
with the corresponding left eigenfunctions of $\mathcal{M}$.
Defining $\delta \mathcal{G}_\alpha=\wh{\mathcal{G}}_\alpha-\mathcal{G}_\alpha,
\,\,\,\,\delta\mathcal{M}^*\equiv \wh{\mathcal{M}}^*-\mathcal{M}^*,$ it
follows from $
\mathcal{M}^*\mathcal{G}_\alpha=-\lambda_\alpha\mathcal{G}_\alpha,
\wh{\mathcal{M}}^*\mathcal{\wh{G}}_\alpha=
-\lambda_\alpha\mathcal{\wh{G}}_\alpha$ that
\be \left(\mathcal{\wh{M}}^*+\lambda_\alpha\right)\delta\mathcal{G}_\alpha
      =-\delta\mathcal{M}^*\mathcal{G}_\alpha. \lb{delG-eq} \ee
A straightforward calculation using (\ref{Mstar-def-I}) and
(\ref{Mstar-def-II}) gives
\begin{eqnarray}
(\delta\mathcal{M}^*)_{k\ell}^{pq}\mathcal{G}_{pq} &=&
\partial_k\left[T^{pq}\left(\partial_q\mathcal{G}_{p\ell}
                        -\partial_\ell\mathcal{G}_{pq}\right)\right]\cr
& &  \,\,\,\,\,\,\,\,
+\partial_\ell\left[T^{pq}\left(\partial_p\mathcal{G}_{kq}
                        -\partial_k\mathcal{G}_{pq}\right)\right]\cr
& &  \,\,\,\,\,\,\,\, +\partial_k\partial_\ell
\left(T^{pq}\mathcal{G}_{pq}\right).
\lb{delM-def} \end{eqnarray}
The solvability condition of eq.(\ref{delG-eq}) is
$\langle\mathcal{\wh{C}}_\alpha,\delta\mathcal{M}^*
\mathcal{G}_\alpha\rangle=0,$ for the right eigenfunction
$\mathcal{\wh{C}}_\alpha$ of
$\mathcal{\wh{M}}$ (which is also the right eigenfunction $\mathcal{C}_\alpha$
of $\mathcal{M}$)
with eigenvalue $\lambda_\alpha.$ This is easily seen to be satisfied using the
definition of
$\delta\mathcal{M}.$ Thus, eq.(\ref{delG-eq}) has a unique solution
$\delta\mathcal{G}_\alpha$
in the subspace orthogonal to $\mathcal{\wh{C}}_\alpha.$ Defining
$\mathcal{\wh{G}}_\alpha
=\mathcal{G}_\alpha+\delta\mathcal{G}_\alpha,$ it follows that
$$ \langle\mathcal{\wh{C}}_\alpha,\mathcal{\wh{G}}_\alpha\rangle=
    \langle\mathcal{\wh{C}}_\alpha,\mathcal{G}_\alpha\rangle
     =\langle\mathcal{C}_\alpha,\mathcal{G}_\alpha\rangle=1. $$
Similar arguments using eq.(\ref{delG-eq}) show that
$\delta\mathcal{M}^*\mathcal{G}_\alpha$
and $\delta\mathcal{G}_\alpha$ are orthogonal to every solenoidal eigenfunction
$\mathcal{\wh{C}}_\beta=\mathcal{C}_\beta$ and, thus, to the entire subspace of
solenoidal
correlation functions.  Hence the new set of eigenfunctions
$\mathcal{\wh{C}}_\alpha,
\mathcal{\wh{G}}_\alpha$ for all $\alpha$ form another biorthogonal set.

There are several simplifications in the isotropic sector of the model. As
shown
in Appendix \ref{BCR}, any function $\delta\mathcal{G}$ in the
isotropic sector which is orthogonal to all solenoidal correlations must be
pure-gauge:
\be \delta\mathcal{G}_{k\ell}=\partial_k\partial_\ell\Lambda. \lb{pure-gauge}
\ee
(For simplicity, we drop here the spectral index $\alpha$ which labels the
eigenvalues and
eigenfunctions.) Finding $\delta\mathcal{G}$ thus reduces to finding $\Lambda.$
Note furthermore that $\mathcal{M}^*\delta\mathcal{G}=0$ when
$\delta\mathcal{G}$
is pure-gauge (ultimately because gauge functions $\lambda$ do not evolve in
the
gauge-choice of eq.(\ref{A-eq-I})).  Thus, the equation for $\delta\mathcal{G}$
becomes
$$ \left( \delta\mathcal{M}^*+\lambda\right)\delta\mathcal{G}=
      -\delta\mathcal{M}^*\mathcal{G}.$$
and, substituting from (\ref{pure-gauge}),
\begin{eqnarray*}
\partial_k\partial_\ell \left(T^{pq}\partial_p\partial_q\Lambda\right)
+\lambda\partial_k\partial_\ell\Lambda &=&
-\partial_k\left[T^{pq}\left(\partial_q\mathcal{G}_{p\ell}
                        -\partial_\ell\mathcal{G}_{pq}\right)\right]\cr
& &  -\partial_\ell\left[T^{pq}\left(\partial_p\mathcal{G}_{kq}
                        -\partial_k\mathcal{G}_{pq}\right)\right]\cr
& &  -\partial_k\partial_\ell \left(T^{pq}\mathcal{G}_{pq}\right).
\end{eqnarray*}
Finally, there must exist in the isotropic sector a scalar function $\Phi(r)$
such that
\be T^{pq}\left(\partial_p\mathcal{G}_{kq}
                        -\partial_k\mathcal{G}_{pq}\right)=\partial_k\Phi,
\lb{Phi-def} \ee
with $\partial_k\Phi(r)=\hat{r}_k\Phi'(r),$ so that the previous equation
becomes
$$ T^{pq}\partial_p\partial_q\Lambda+\lambda\Lambda
      =-\left[2\Phi+T^{pq}\mathcal{G}_{pq}\right]. $$
Substituting the isotropic form
$$ T^{pq}(\br) = T_L(r)\hat{r}^p\hat{r}^q +
T_N(r)\left(\delta^{pq}-\hat{r}^p\hat{r}^q\right)
     + T_H(r) \epsilon^{pqm}\hat{r}_m, $$
and the similar expression for $\mathcal{G}$ into (\ref{Phi-def}) gives, after
some computation,
$\Phi'(r) = 2 (T_N \Psi_H - T_H\Psi_N). $ The resulting equation to be solved
for $\Lambda$ is
\be
T_L\partial_r^2\Lambda + 2T_N\frac{1}{r}\partial_r\Lambda
+\lambda \Lambda = -\left[2\Phi+T^{pq}\mathcal{G}_{pq}\right], \lb{Lam-eq} \ee
with
$$ \Phi(r) = -2 \int_r^\infty
d\rho\,[T_N(\rho)\Psi_H(\rho)-T_H(\rho)\Psi_N(\rho)] $$
(so that $\Phi(+\infty)=0$) and
$$ T^{pq}G_{pq}=T_LG_L+2T_NG_N+2T_HG_H. $$

Alternatively, the equation (\ref{Lam-eq}) may be written using BCR quantities,
with $T_L=b,\,\,\,T_N=a/\sqrt{2}$
$$ \Phi' = \frac{aW_2-cW_3}{r}-\sqrt{2}c \Psi_L, $$
$$\Psi_L(r)=\sqrt{2}\int_0^r d\rho\,\frac{W_3(\rho)}{\rho^2},$$
and
$$ TG\equiv T^{pq}G_{pq}= \frac{a G_1 + bG_2 + cG_3}{r}. $$

\subsection{Numerical Results and Physical Discussion} \lb{numerical}

As a particular case of physical interest we shall consider an incompressible,
non-helical velocity field with $a=\sqrt{2}(b+\frac{1}{2}rb')$ and $c=0.$ To
model
an infinite-Reynolds-number inertial-range, we take, for $0<\xi<2,$
$$ b(r)=2\eta + 2D_1 r^\xi.  $$
We non-dimensionalize the equations of the KK model using
$\ell_\eta=(\eta/D_1)^{1/\xi}$
for space and $\tau_\eta=\ell_\eta^2/\eta$ for time. The Sturm-Liouville
problem
is then given by (\ref{SL-eq}), or
\begin{eqnarray}
  &&\,\,\,\,\,\,\,\,\,\,\,\,-\partial_r(p(r)\partial_rW_2)+q(r)W_2 = \lambda
W_2, \cr
  && p(r)=2(1+r^\xi), \,\,\,\,q(r)=
{{4}\over{r^2}}-{{2(\xi+2)(\xi-1)}\over{r^{2-\xi}}}.
  \,\,\,\,\,\,\,\,\,
\lb{SL-eq-p} \end{eqnarray}
Endpoints $r=0,\infty$ are singular, non-oscillatory and limit-point.  The
Friedrichs boundary
conditions to select the principal solution are $W_2(0)=W_2(\infty)=0.$ E.g.
see \cite{Zettl05}.
The Sturm-Liouville operator with $q(r)$ above is obviously non-negative for
$\xi<1,$ so
dynamo effect can exist only for $\xi>1.$ In the latter case, there is both
point spectrum and, above
some threshold value, $\lambda>\lambda_c,$ continuous spectrum. The lowest
eigenvalue is negative,
implying dynamo effect.  To give the closest correspondence with real
hydrodynamic
turbulence,  we take the Richardson value $\xi=4/3$ in our work here. See
\cite{Vincenzi02}
for a numerical study of general values $1<\xi<2.$

We obtain the solution of (\ref{SL-eq-p})
using the {\tt MATSLISE} software for ${\tt MATLAB}$
\cite{Ledouxetal05,Ledoux07}.
Briefly, this package solves regular Sturm-Liouville eigenvalue problems by
first converting
them through the so-called Liouville transformation $W_2=\Psi/[p(r)]^{1/4}$ and
$x=\int_0^r
\frac{dr'}{\sqrt{p(r')}}$into a corresponding Schr\"odinger equation for
$\Psi(x).$ The latter is
then solved by a high-order Constant Perturbation Method (CPM) algorithm, which
generates
its own numerical grid of points $r_i,$ $i=1,2,...,L.$ To treat our singular
problem on
an infinite interval, we truncate to a finite interval $[0,b]$ with the
condition $W_2(b)=0$
and then take successively larger $b$ values, as discussed in \cite{Ledoux07},
Section 6.3.1.
We have considered successively $b=200,...,700,800,$ with convergence of
solutions for $r<500.$  The ground-state eigenvalue found by this method is
$\lambda\doteq -0.1927,$ in good agreement with \cite{Vincenzi02}, Fig. 5,
and the corresponding eigenfunction $W_2(r)$ is plotted in Fig.~1. The
eigenfunction is normalized so that
\be \int_0^\infty dr\,W_2^2(r)=1. \lb{W-norm} \ee

\begin{figure}
\centerline{\includegraphics[width=9cm,height=7cm]{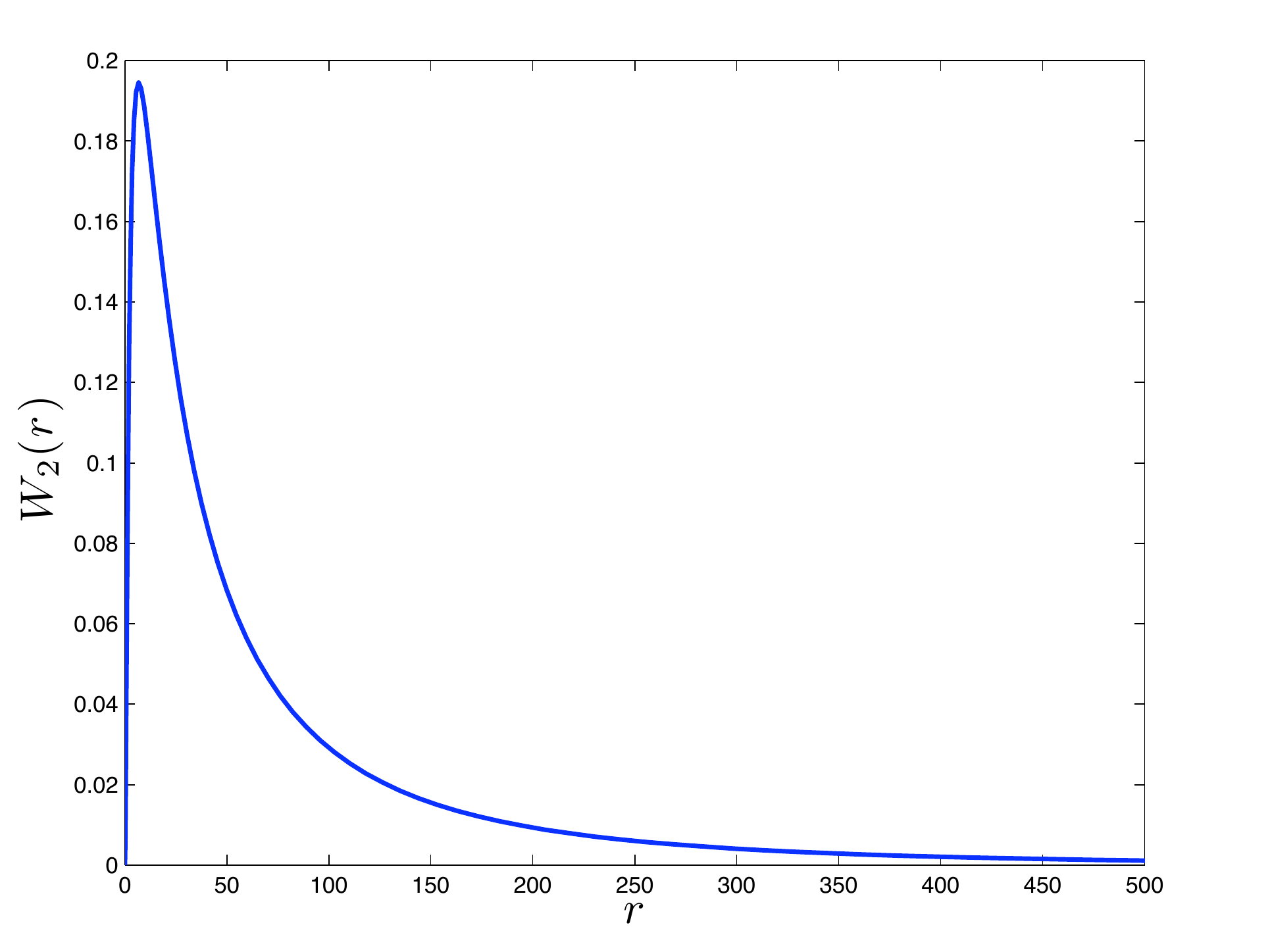}}
\caption{Plot of ground-state eigenfunction $W_2(r)$ versus $r.$ All quantities
in this and following figures have been non-dimensionalized with resistive
units as discussed in the text.}
\end{figure}\label{eigfun-fig}

\begin{figure}
\centerline{\includegraphics[width=9cm,height=7cm]{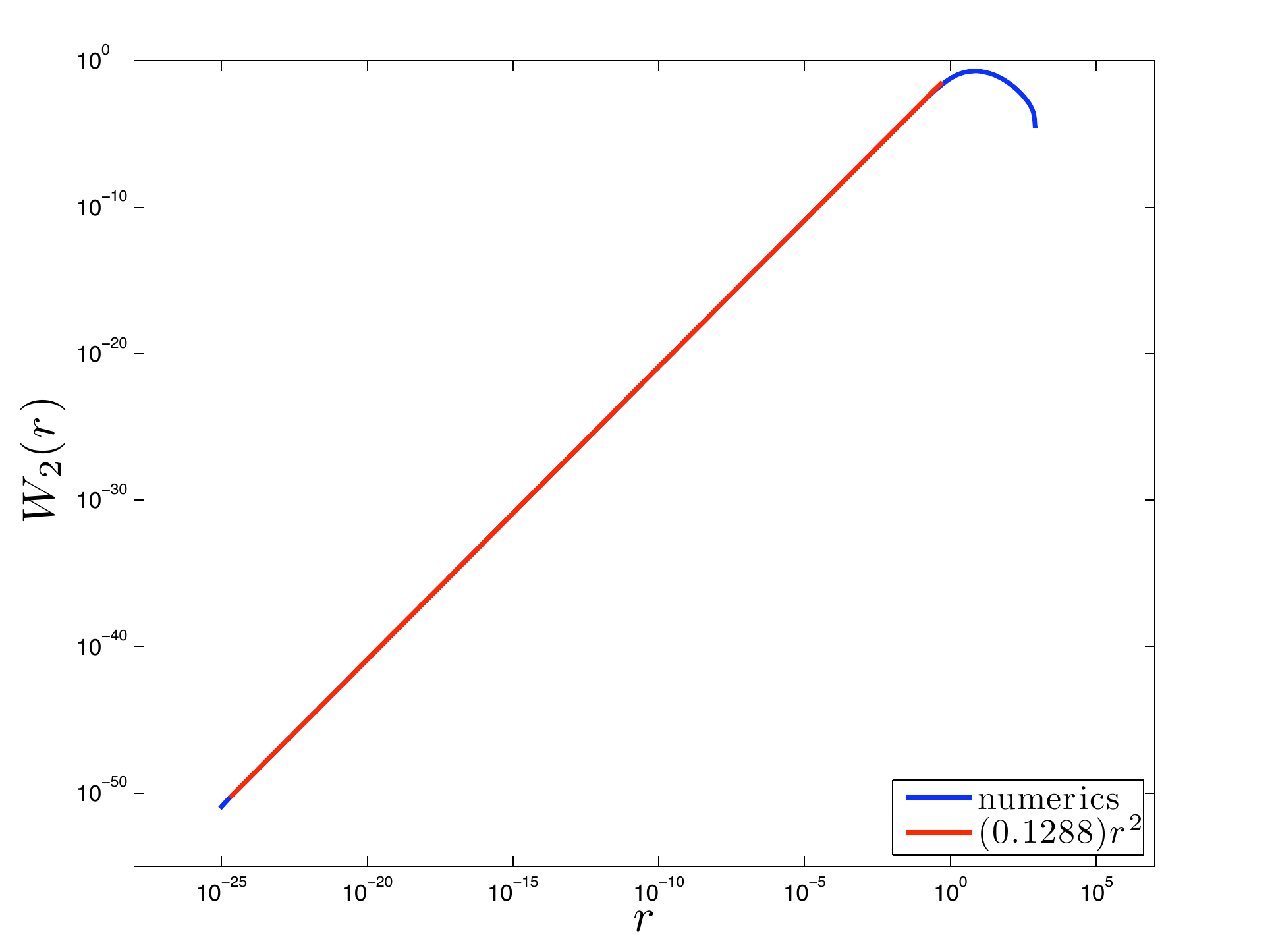}}
\caption{Small-$r$ asymptotics of ground-state eigenfunction.
In \textcolor{blue}{blue} is a log-log plot of $W_2(r)$ versus $r$ and
in \textcolor{red}{red} the quadratic fit $(0.1288)r^2.$}
\end{figure}\label{Wsmallr}

\begin{figure}
\centerline{\includegraphics[width=9cm,height=7cm]{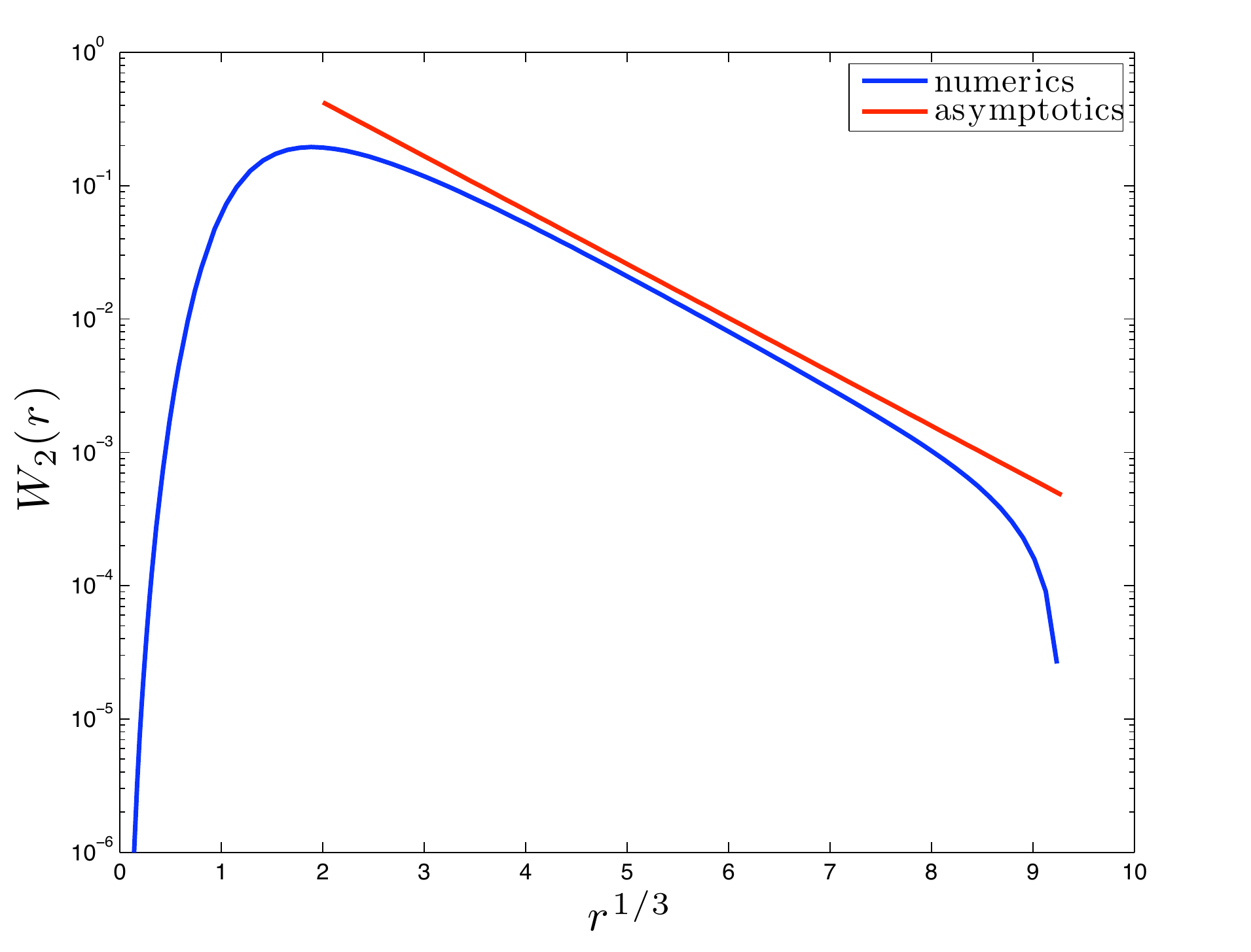}}
\caption{Large-$r$ asymptotics of ground-state eigenfunction.
In \textcolor{blue}{blue} is a log-linear plot of $W_2(r)$ versus $r^{1/3}$ and
in \textcolor{red}{red} a straight line with slope from (\ref{W2-large}) for
$\xi=4/3$.}
\end{figure}\label{Wlarger}

Asymptotics at small and large $r$ are known for the ground-state
eigenfunction.
The small-$r$ behavior was obtained by \cite{BoldyrevCattaneo04}, who noted
\be W_2(r) = A r^2 + B r^{2+\xi} + o(r^{2+\xi}),\,\,\,\,\,r\ll 1. \lb{W2-small}
\ee
Indeed, direct substitution of this ansatz shows that the Sturm-Liouville
eigenvalue
equation (\ref{SL-eq-p}) is then satisfied up to terms $o(r^\xi),$ if and only
if $B=-A.$
The large-$r$ behavior from a
WKB asymptotic analysis is \cite{Vincenzi02,BoldyrevCattaneo04}
\be W_2(r) \sim \exp\left[-{{\sqrt{-2\lambda}\,
r^{(2-\xi)/2}}\over{2-\xi}}\right],\,\,\,\,\,r\gg 1
\lb{W2-large} \ee
up to a power-law prefactor.  This  corresponds to the dominant balance $2r^\xi
W_2''
\doteq -\lambda W_2$ in the Sturm-Liouville eq.(\ref{SL-eq-p}) at large-$r$.
Both the small-$r$ and large-$r$ asymptotic behaviors predicted analytically
have been verified in our numerical solution. As shown in Fig.~2,
the leading-order $r^2$ behavior in (\ref{W2-small}) is verified over about 24
orders of magnitude with the value $A\doteq 0.1288$ obtained by taking the
small-$r$ limit of the numerical results for $W_2/r^2.$ Although we shall not
show it here, we have furthermore verified the subleading $r^{10/3}$ term in
(\ref{W2-small}) with the same value of $A,$ over about 16 orders of magnitude.
We also verify the stretched-exponential decay (\ref{W2-large}) at large-$r$,
as shown by the log-linear plot of $W_2(r)$ vs. $r^{1/3}$ in Fig.~3. The red
line shows the slope predicted by (\ref{W2-large}) with $\lambda=-0.1927.$
Our numerical evaluation of $W_2(r)$ is in very good agreement with all
known analytical results for the dynamo eigenfunction.

Magnetic and vector-potential correlations in the dynamo growth mode are
obtained from
$W_2$ via (\ref{R-eigfun}), (\ref{L-eigfun}) and plotted in Figs.~4 and 5 using
the traditional
longitudinal and transverse functions. The normalizations are those which
follow from
(\ref{W-norm}).  One obvious feature is the much longer range of the
vector-potential
correlations as compared with the magnetic-field correlations.
\begin{figure}
\centerline{\includegraphics[width=9cm,height=7cm]{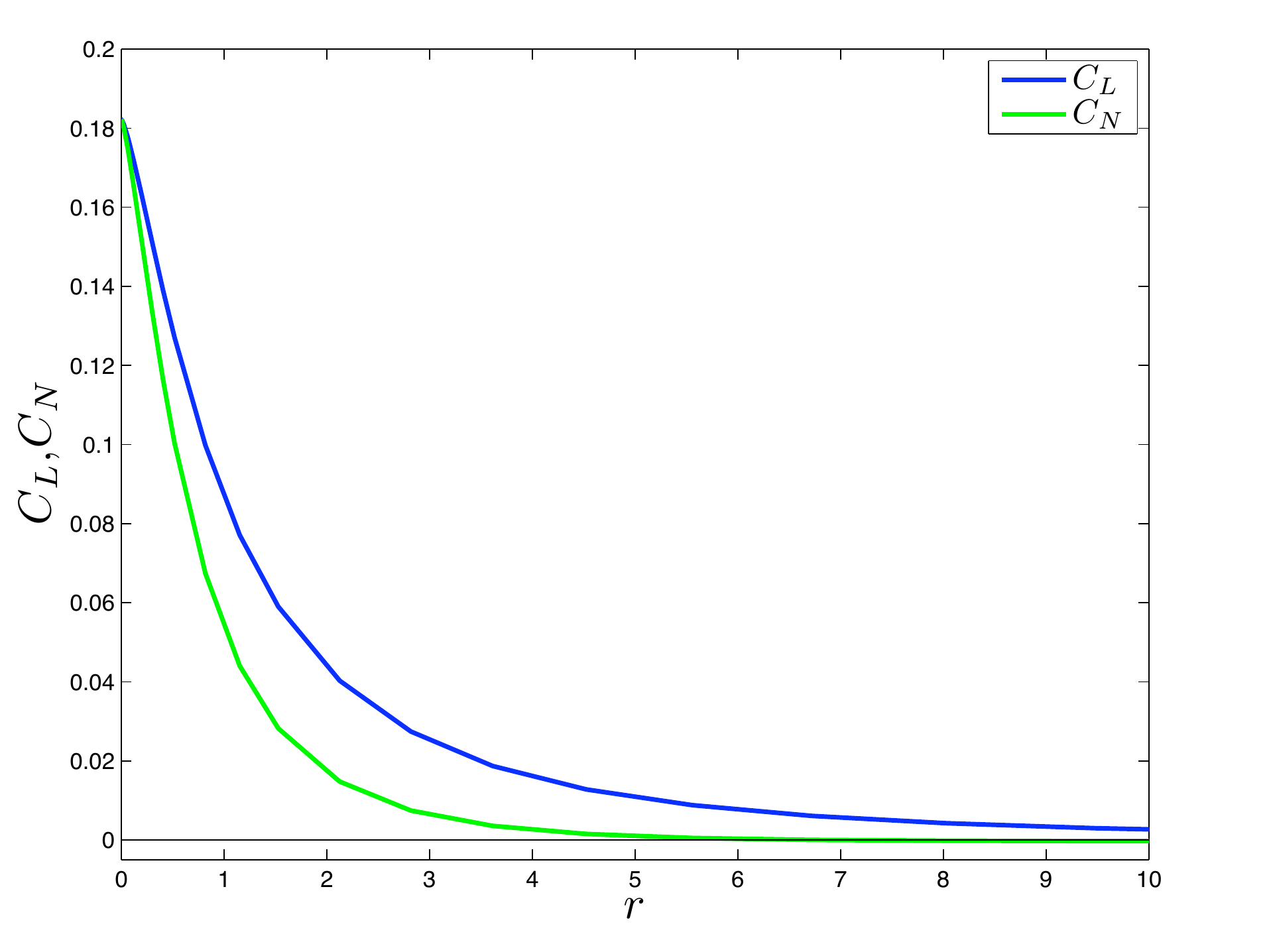}}
\label{Bcorr-fig}
\caption{Plot of magnetic field correlation functions, longitudinal $C_L(r)$ in
\textcolor{blue}{blue} and transverse $C_N(r)$ in \textcolor{green}{green}
versus $r.$\\}
\end{figure}
\begin{figure}
\centerline{\includegraphics[width=9cm,height=7cm]{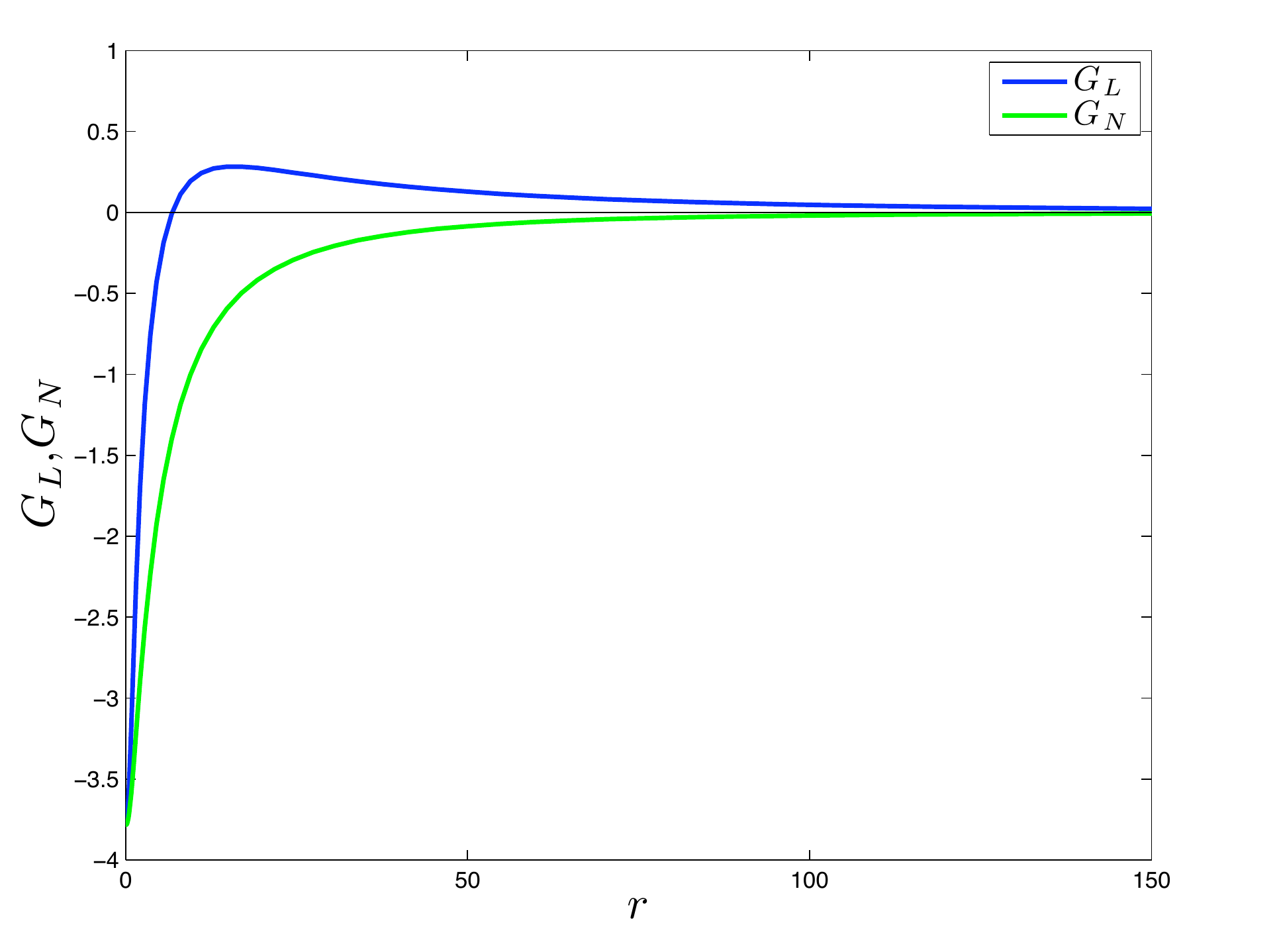}}
\label{Acorr-fig}
\caption{Plot of magnetic vector-potential correlation functions, longitudinal
$G_L(r)$
in \textcolor{blue}{blue} and transverse $G_N(r)$ in \textcolor{green}{green}
versus $r.$}
\end{figure}
The small-$r$ behavior is obtained from (\ref{W2-small}) to be
\be C_L \sim \sqrt{2}A [1-r^{\xi}+o(r^{\xi})], \lb{CL-small} \ee
\be C_N \sim \sqrt{2}A[1-(1+\frac{\xi}{2})r^{\xi}+o(r^{\xi})], \lb{CN-small}
\ee
and
\be G_L \sim  G_N  \sim  \frac{4\sqrt{2}A}{\lambda}[1+o(r^\xi)]. \lb{GLN-small}
\ee
Note that the contributions of order $r^{\xi}$ cancel in the latter two
functions,
signalling their greater smoothness. The large-$r$ behavior is found, with
(\ref{W2-large}), to be
\be
C_1 \sim -\sqrt{\frac{-\lambda}{2}} r^{-\xi/2}W_2,\,\,\,\,C_2 \sim
{{\sqrt{2}}\over{r}} W_2
\lb{C-large} \ee
\be G_1 \sim  \sqrt{\frac{2}{-\lambda}} r^{\xi/2} W_2,
 \,\,\,\, G_2  \sim \frac{2+\xi}{\sqrt{-\lambda}} r^{\xi/2} W_2. \lb{G-large}
\ee
The signs implied for $r\gg 1$ are $C_N<0,\,\,C_L>0$ and $G_N,\,\,G_L>0.$
Although it is difficult to see clearly in Fig.~4, $C_N(r)<0$ in our numerical
solution for $r>6.85.$ The existence of negative tails in $C_N$ was noted
some time ago  \cite{RuzmaikinSokolov81,Novikovetal83} to be a consequence
of the solenoidal character of the magnetic field (see also just below).

 The physical-space behaviors discussed above imply corresponding results for
the
 magnetic energy spectrum of the dynamo mode. At low wavenumbers
 \be E(k)\sim A' k^4, \,\,\,\,\,\,\,\, k\ll k_\eta  \lb{energy-low} \ee
 with  $k_\eta=2\pi/\ell_\eta$ and $A'$ some constant numerically proportional
to $A.$
 To see this, note that the stretched-exponential decay (\ref{C-large}) implies
 that $E(k)$ is analytic at low-$k$ and can be expanded as a convergent
 power-series in $k^2.$ The leading term proportional to $k^2$ vanishes,
however,
 since the integral $\int d^3r\, \mathcal{C}^{ij}(\br)=0.$  By isotropy, this
is equivalent
 to the vanishing of the integral
$\int_0^\infty dr\,r^2 C_T,$ where  $C_T={{\sqrt{2}C_1+C_2}\over{r}}$ is the
trace
$C_T=C_{ii}.$ Using (\ref{R-eigfun}),
$$
r^2C_T=\sqrt{2}\left(r\partial_rW_2+W_2\right)=\sqrt{2}\partial_r\left(rW_2\right), $$
so that
$$ \int_0^\infty dr\,r^2 C_T(r)=\sqrt{2} \lim_{r\rightarrow \infty} rW_2(r)=0.
$$
Note that this result requires the existence of negative tails in $C_T$ at
large $r$
 \cite{RuzmaikinSokolov81,Novikovetal83}. The $k^4$ spectrum in
(\ref{energy-low})
was explained long ago by Kraichnan and Nagarajan \cite{KraichnanNagarajan67}
as due to the dipole magnetic  field that results from ``irregularly twisted
and elongated
current loops whose transverse dimension is $\sim k_m^{-1}$ [ our
$k_\eta^{-1}$].''

The spectrum at high wavenumbers is \cite{BoldyrevCattaneo04}
\be E(k)\sim C k^{-(1+\xi)},  \,\,\,\,\,\,\,\, k\gg k_\eta. \lb{energy-high}
\ee
This follows mathematically by Fourier transforming the singular terms $\propto
r^\xi$
in (\ref{CL-small}),(\ref{CN-small}). The result (\ref{energy-high}) is the
exact analogue
for the Kazantsev model of the Golitsyn-Moffatt $k^{-11/3}$ spectrum
\cite{Golitsyn60,Moffatt61}.  The dominant balance in the induction equation
for
length-scales $\ell\ll\ell_\eta$ is
$$ \partial_t\bB -\eta\triangle\bB = \bB_\eta\bdot\grad\bu, $$
with $\bB_\eta$ the magnetic field at scale $\ell_\eta.$ The time-derivative
must be included
because of the rapid change in time of the velocity field. Solving for the
statistical steady-state
of the above linear Langevin equation leads to $E(k)\sim \frac{\langle
B^2\rangle}{\eta} E_u(k),$
where $E_u(k)$ is the energy spectrum  of the velocity field, and not to
$E(k)\sim
\frac{\langle B^2\rangle}{ \eta^2k^2} E_u(k),$ as in the original argument of
Golitsyn.
Our discussion here exactly parallels that of Frisch and Wirth
\cite{FrischWirth96}
for the analogous passive scalar problem. Note finally that the energy spectrum
in the
Kazantsev model, both in its low-$k$ and high-$k$ behaviors, is thus close to
that
argued for kinematic dynamo in inertial-range hydrodynamic turbulence by
Kraichnan and Nagarajan \cite{KraichnanNagarajan67}.

\begin{figure}
\centerline{\includegraphics[width=9cm,height=7cm]{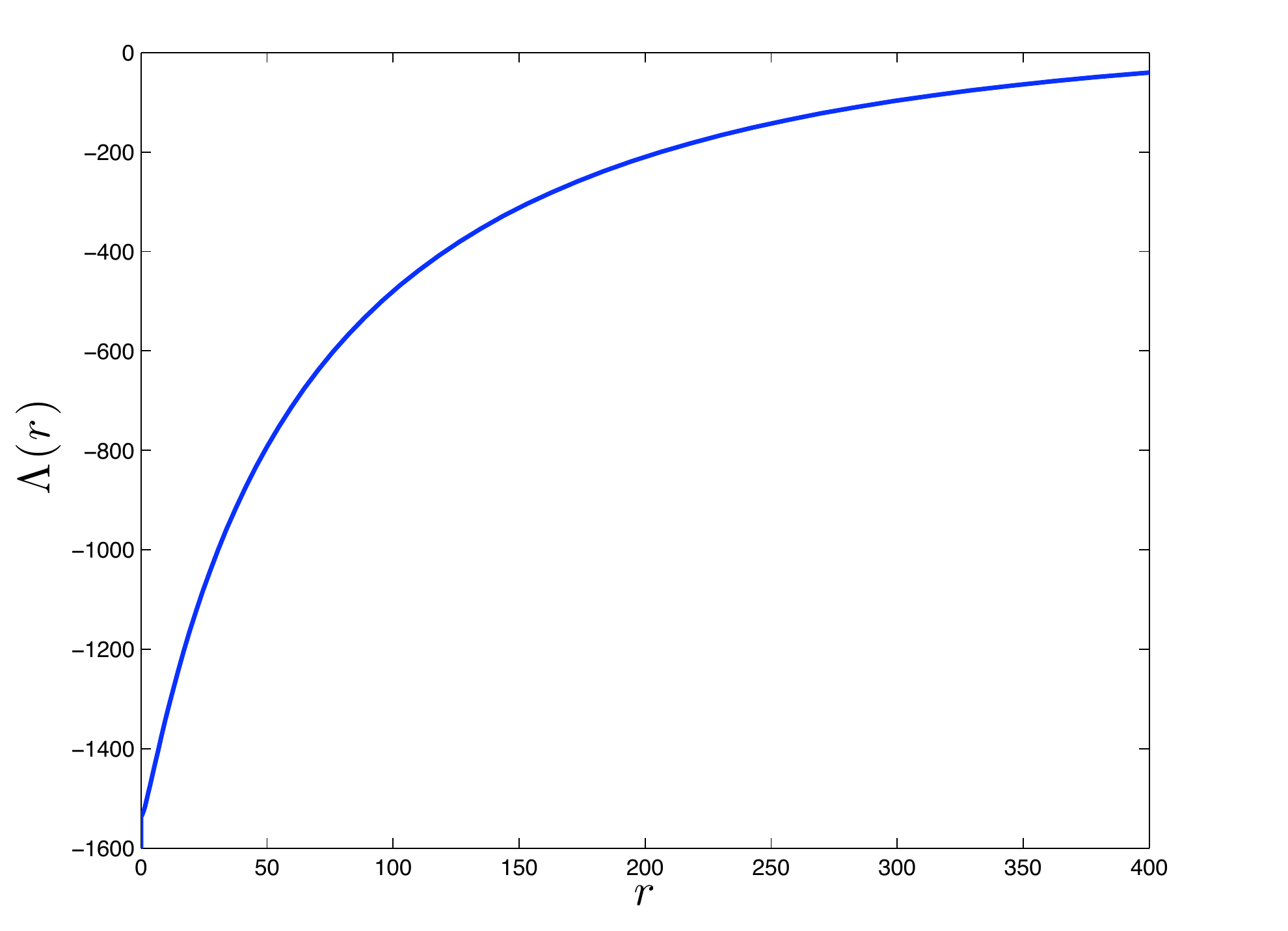}}
\label{Lambda-fig}
\caption{Plot of gauge-change function $\Lambda(r)$ versus $r$.}
\end{figure}

\begin{figure}
\centerline{\includegraphics[width=9cm,height=7cm]{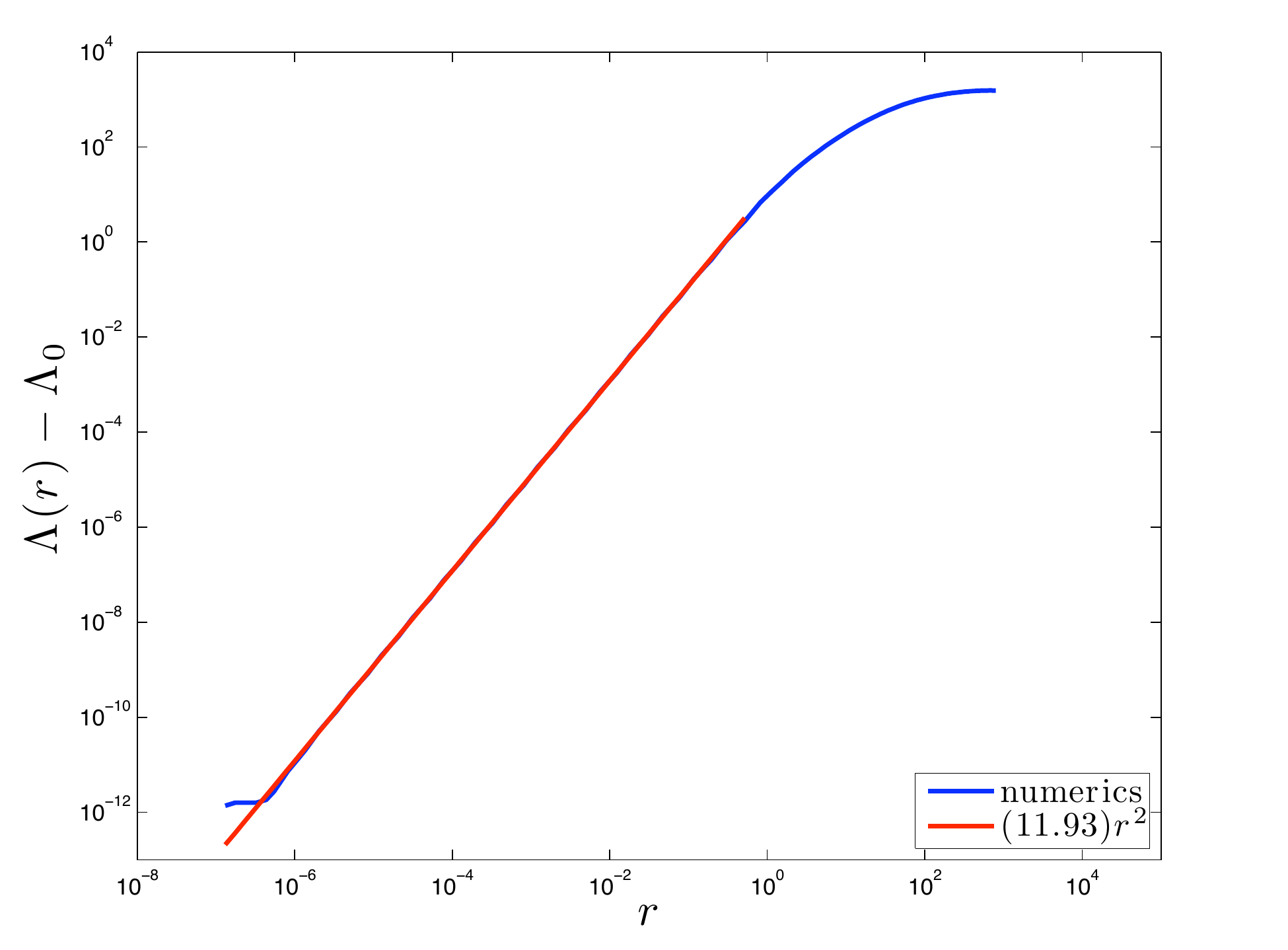}}
\label{Lam-small}
\caption{Small-$r$ asymptotics of gauge-change function. In
\textcolor{blue}{blue}
is a log-log plot of $\Lambda(r)-\Lambda_0$ versus $r$ and
in \textcolor{red}{red} the quadratic fit $(11.93)r^2.$}
\end{figure}

\begin{figure}
\centerline{\includegraphics[width=9cm,height=7cm]{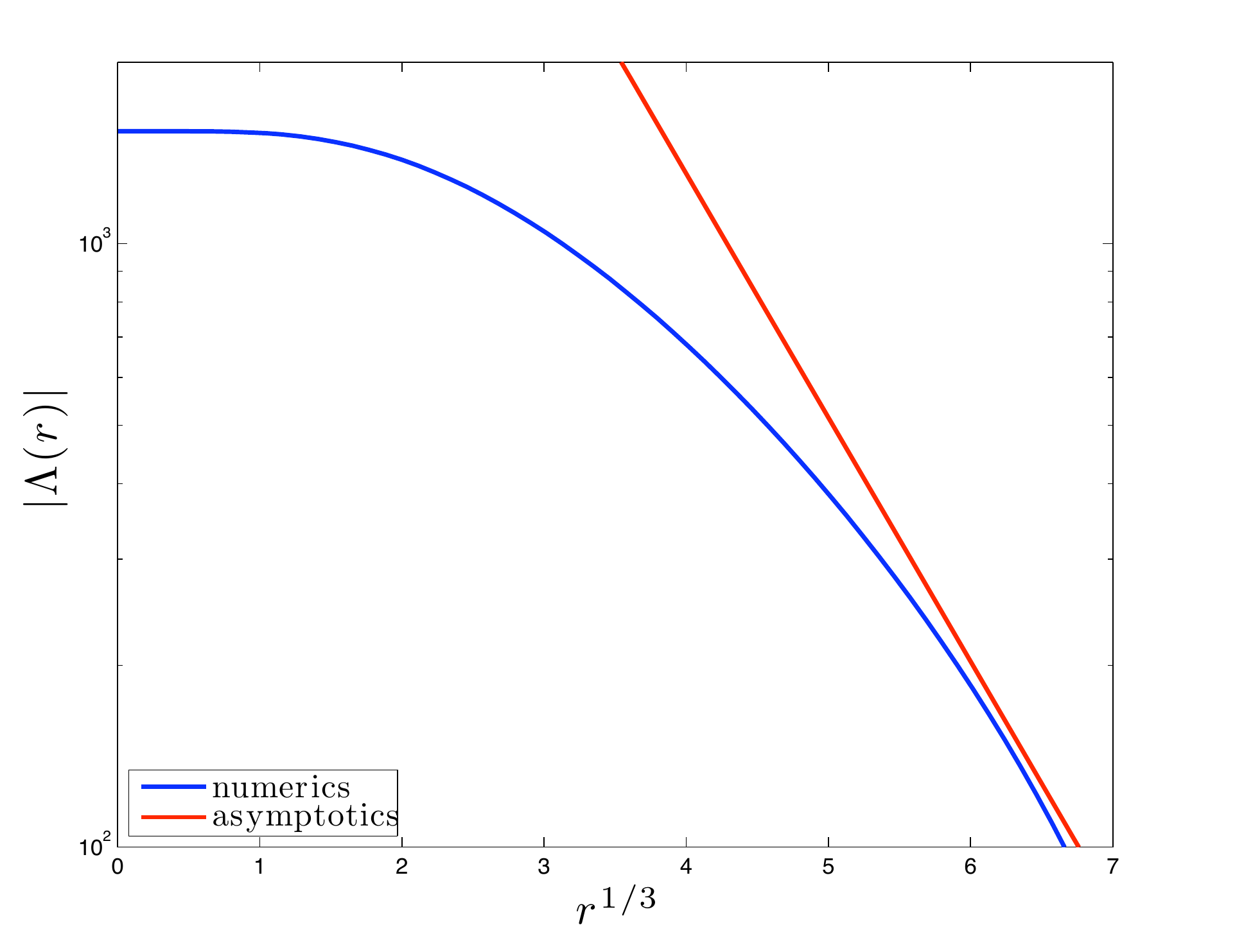}}
\label{Lam-large}
\caption{Large-$r$ asymptotics of gauge-change function. In
\textcolor{blue}{blue}
is a log-linear plot of $|\Lambda(r)|$ versus $r^{1/3}$ and in
\textcolor{red}{red} a straight
line with slope from (\ref{Lambda-large}) for $\xi=4/3$.}
\end{figure}

We shall now turn to the evaluation of the gauge-II correlations. For this
purpose, we
must determine the gauge-change function $\Lambda(r)$ which solves
eq.(\ref{Lam-eq}):
\be r^{-2}(r^2b\Lambda')'+\lambda\Lambda=-F \lb{Lam-eq2} \ee
with $F=2\Phi+TG.$ We have solved this equation numerically, by the procedure
described
in Appendix \ref{numerics}. The result is plotted in Fig.~6.

At small $r$, the solution $\Lambda=\Lambda_{reg}+\Lambda_{sing}$ is the sum
of a regular part
\be \Lambda_{reg}\sim \Lambda_0 +\Lambda_1r^2+o(r^2) \lb{Lam-reg} \ee
and a singular part
\be \Lambda_{sing}\sim \Lambda_s r^{2+\xi}+o(r^{2+\xi}). \lb{Lam-sing} \ee
This is verified by substituting into eq.(\ref{Lam-eq2}), using the similar
expansion for $F$
$$ F\sim F_0 + F_1 r^\xi $$
which follows from
$$ \Phi = \Phi_0+\int_0^r d\rho\, \frac{aW_2}{\rho} = \Phi_0 +O(r^2) $$
with $\Phi_0=-\int_0^\infty dr\, \frac{aW_2}{r}<0$ and
$$ TG= \frac{8\sqrt{2}A}{\lambda}\left[3+(3+\xi)r^\xi+o(r^\xi)\right]. $$
Thus, $F_0=2\Phi_0+\frac{24\sqrt{2}A}{\lambda}<0$ and $F_1=
\frac{8\sqrt{2}A}{\lambda}(3+\xi)<0.$ Then it is straightforward to obtain
\be \lambda\Lambda_0+12\Lambda_1=-F_0 \lb{F0-eq} \ee
\be 2\left[2\Lambda_1 +(2+\xi)\Lambda_s\right](3+\xi)= -F_1. \lb{F1-eq} \ee
These results have been verified in our numerical solution, with specific
values
of the constants
$$ \Lambda_0 = -1535.5, \,\,\Lambda_1 =11.9299, \,\,
      \Lambda_s =-6.0232. $$
See Appendix \ref{numerics}, and Fig.~7 for a comparison of $\Lambda(r)$
and $\Lambda_{reg}(r)$ at small $r.$

At large-$r$ we expect $\Lambda(r)\sim \exp\left[-{{\sqrt{-2\lambda}\,
r^{(2-\xi)/2}}\over{2-\xi}}\right]$
up to power-law prefactors.  Rewriting the eq.(\ref{Lam-eq2}) for $\Lambda$ as
$$ b\Lambda'' + r^{-2}(r^2b)'\Lambda'+\lambda \Lambda= -F $$
one finds that the terms $b\Lambda'',\,\,\lambda\Lambda$ cancel to leading
order.
For $r\gg 1.$
$$ TG \sim \frac{2(2+\xi)}{\sqrt{-\lambda}}r^{3\xi/2-1}W_2, $$
and $\Phi'=aW_2/r\sim \sqrt{2}(2+\xi)r^{\xi-1}W_2,$ so that
$$ \Phi \sim -\frac{2(2+\xi)}{\sqrt{-\lambda}}r^{3\xi/2-1}W_2. $$
It follows that
$$ F = 2\Phi+TG\sim \Phi. $$
The dominant balance of $r^{-2}(r^2b)'\Lambda'$ and $-F$ gives
\be \Lambda \sim \frac{\sqrt{2}}{\lambda} r^\xi W_2, \,\,\,\,\,\,\, r\gg 1.
\lb{Lambda-large} \ee
This asymptotics is verified in our numerical solution. See Fig.~8 for a
log-linear
plot of $|\Lambda(r)|$ vs. $r^{1/3},$ with the red line having the slope
predicted by
(\ref{Lambda-large}),(\ref{W2-large}).

{}From the function $\Lambda$ we obtain $\delta\mathcal{G}$ via
eq.(\ref{pure-gauge}).
We plot in Fig.~9 the longitudinal and transverse components $\delta G_L$ and
$\delta G_N$
obtained from (\ref{gradient}). It is interesting that these functions show
almost exactly opposite
behaviors of algebraic signs compared with $G_L,\,\,G_N$ in Fig.~5.  Both
$\delta G_L$ and
$\delta G_N$ are proportional to $r^\xi$ at small $r$ because of the
$\Lambda_s$ contribution
and are stretched-exponentials
\be \delta G_1 \sim \sqrt{\frac{2}{-\lambda}} r^{\xi/2} W_2, \,\,\,\,
       \delta G_2 \sim -\frac{r}{\sqrt{2}}W_2 \lb{delG-large} \ee
at large $r,$ so that $\delta G_N>0,\,\,\delta G_L<0$ for $r\gg 1.$

\begin{figure}
\centerline{\includegraphics[width=9cm,height=7cm]{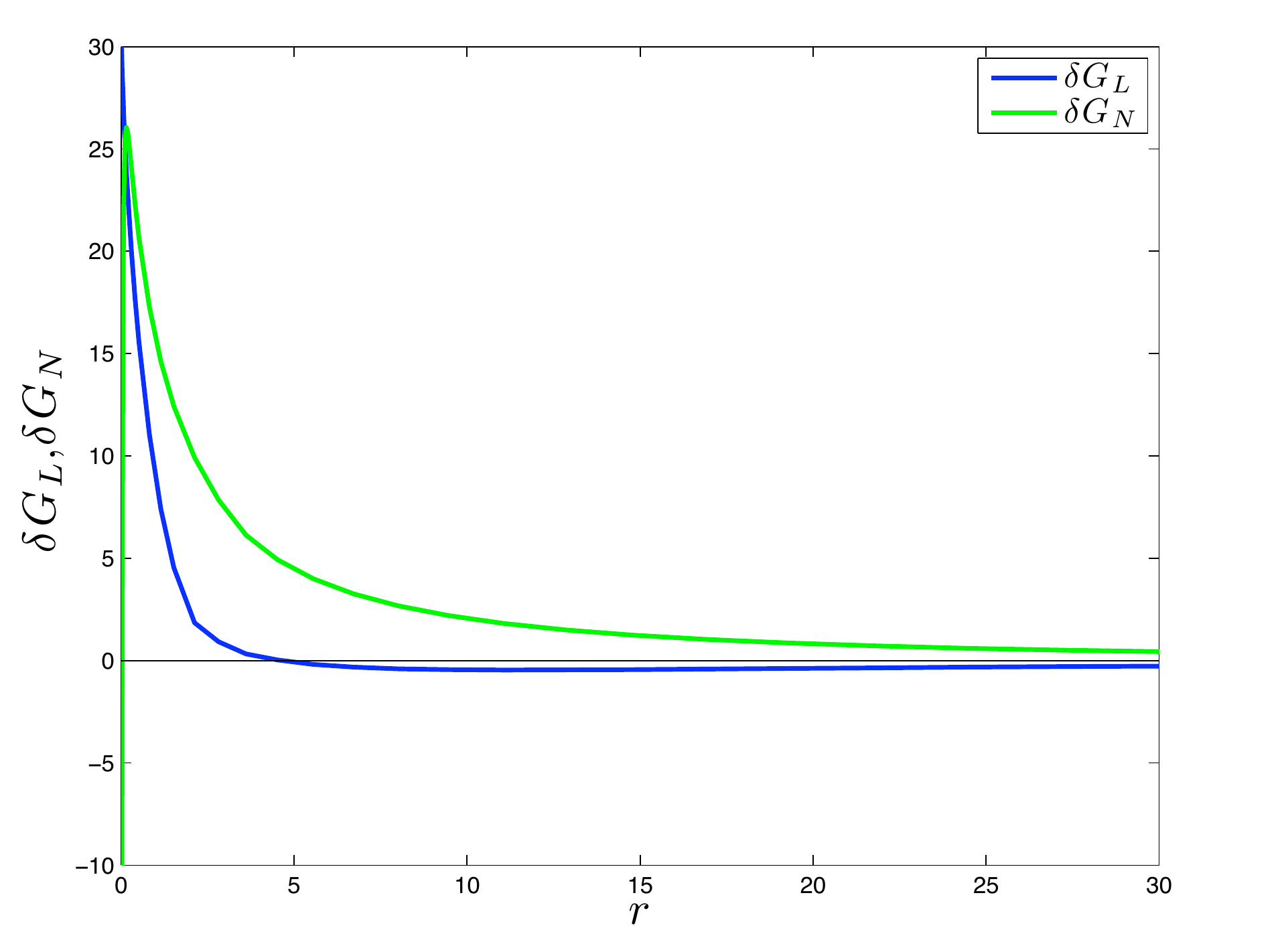}}
\label{deltaG-fig}
\caption{Plot of change in vector-potential correlation functions, longitudinal
$\delta G_L(r)$ in \textcolor{blue}{blue} and transverse $\delta G_N(r)$ in
\textcolor{green}{green}, versus $r.$}
\end{figure}

\begin{figure}
\centerline{\includegraphics[width=9cm,height=7cm]{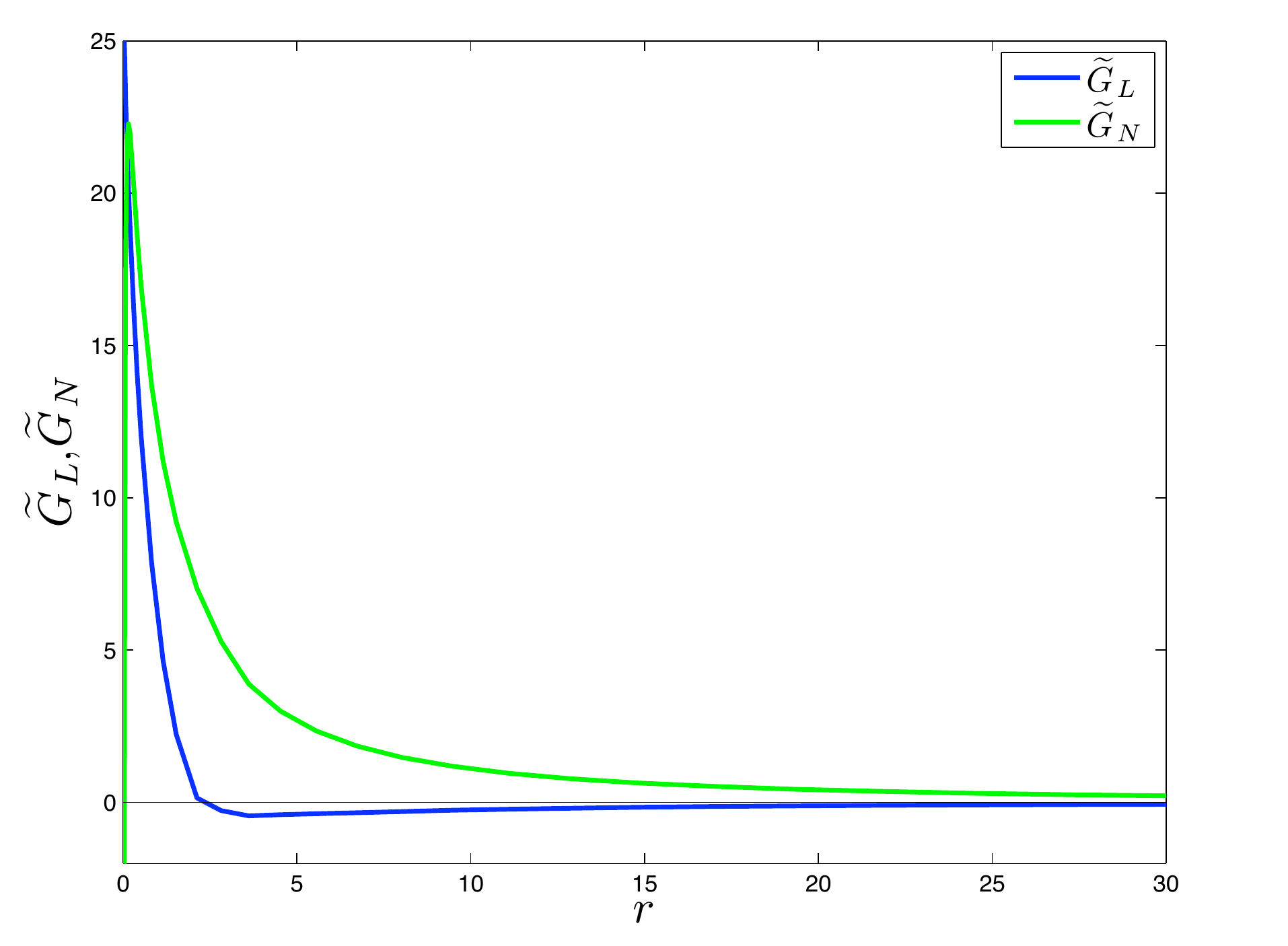}}
\label{tildeG-fig}
\caption{Plot of line-vector correlation functions, longitudinal $\wh{G}_L(r)$
in \textcolor{blue}{blue} and transverse $\wh{G}_N(r)$ in
\textcolor{green}{green},
versus $r.$}
\end{figure}

Finally, we obtain $\mathcal{\wh{G}}=\mathcal{G}+\delta\mathcal{G}.$ Plotted
in Fig.~10 are $\wh{G}_L$, $\wh{G}_N.$ These are the central results of this
paper,
because of their relation to $\mathcal{R}$ in eq.(\ref{R-longT}). We can thus
interpret
these functions as the correlations in the dynamo growth phase of vector
line-elements
carried to the same final point that were initially separated by distance $r,$
with directions
longitudinal and transverse to the separation vector, respectively. We shall
discuss
their physical interpretation below; first we consider the asymptotic behaviors
for
small-$r$ and large-$r$. Combining results for $\mathcal{G}$ and
$\delta\mathcal{G}$
at $r\ll 1$ gives
\be \wh{G}_L \sim \left( \frac{4\sqrt{2}A}{\lambda}+2\Lambda_1\right)
   +(2+\xi)(1+\xi)\Lambda_s r^{\xi} \lb{G-small-II} \ee
and
\be \wh{G}_N \sim \left(\frac{4\sqrt{2}A}{\lambda}+2\Lambda_1\right)
    +(2+\xi)\Lambda_s r^{\xi}.  \lb{GN-small-II} \ee
It is interesting to note that $\mathcal{\wh{G}},$ unlike $\mathcal{G},$
contains
terms $\propto r^\xi$ and is singular at small $r$. The reason for this is that
the
eq.(\ref{A-eq-II}) for $\wh{\bA}$ contains the velocity-gradient $\grad\bu,$
whereas
the eq.(\ref{A-eq-I}) for $\bA$ contains only $\bu$ itself. Finally, comparing
the
large-$r$ asymptotics in (\ref{G-large}) with that in (\ref{delG-large}) gives
for
$r\gg 1$
\be  \wh{G}_1 \sim 2\sqrt{\frac{2}{-\lambda}} r^{\xi/2} W_2, \,\,\,\,
     \wh{G}_2 \sim -\frac{r}{\sqrt{2}}W_2, \lb{G-large-II} \ee
so that $\wh{G}_N>0,\,\,\wh{G}_L<0$ for $r\gg 1.$

The results in Fig.~10 quantify the significance of magnetic field-line
stochasticity
for the small-$Pr_m$ turbulent dynamo. The plotted correlations represent the
contribution to magnetic energy at a given point produced by field vectors that
arrive
from points separated by distance $r$ initially.  As one can see, the
contribution is quite
diffuse (in units of the resistive length $\ell_\eta$) with fat,
stretched-exponential tails.
Both correlations are positive for small separations $r <3\ell_\eta.$
Particularly interesting is the long negative tail for the longitudinal
correlation
$\wh{G}_L,$ which represents an ``anti-dynamo effect'' that suppresses the
turbulent growth of magnetic field energy.

\begin{figure}
\centerline{\includegraphics[width=9cm,height=7cm]{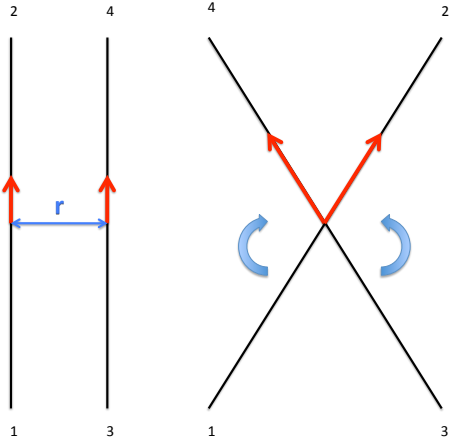}}
\label{Transverse-fig}
\caption{Contribution of transverse seed field vectors to a positive
line-vector
correlation. Two transverse field-vectors on parallel lines at the initial time
(left), when brought to the same point by a fluid motion, are positively
correlated at first contact of the lines (right).}
\end{figure}

\begin{figure}
\centerline{\includegraphics[width=9cm,height=7cm]{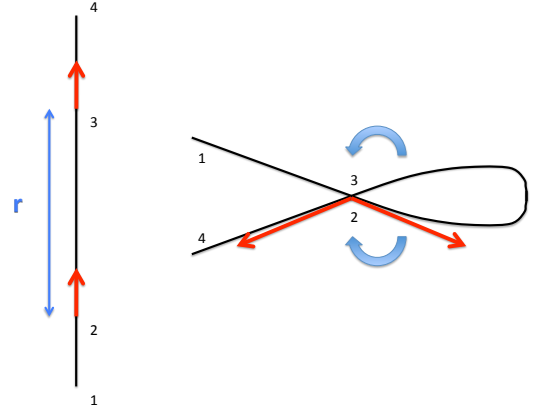}}
\label{Longitude-fig}
\caption{Contribution of longitudinal seed field vectors to a negative
line-vector
correlation. Two longitudinal field-vectors on the same line---or two adjacent
lines---
at the initial time (left), when brought to the same point by a fluid motion,
are
negatively correlated at first contact of the lines (right).}
\end{figure}

It is easy to understand the positive signs of $\wh{G}_L,\,\,\wh{G}_N$ at small
$r$
values. For $r\lesssim \ell_\eta$ the Brownian motion term dominates in the
equation
which follows from (\ref{x-def}) for the relative motion
$\br(t)=\bx_2(t)-\bx_1(t)$ of
a pair of points. Since the Brownian motion corresponds to a simple
translation of the lines, with no rotation, the line-vectors which start
parallel
(either longitudinal or transverse) at separation $r\lesssim \ell_\eta$ will
tend to
remain nearly parallel when they intersect. Hence, they arrive at the
final point positively correlated.

The signs of the tails of  $\wh{G}_L,\,\,\wh{G}_N$ for large $r$ are less easy
to
understand, because turbulent advection effects become important for
$r\gtrsim\ell_\eta.$
We suggest here a heuristic explanation. In the left half of Fig.~11 we show a
pair of
parallel field-lines separated by $\br$ with their initially transverse
line-vectors
indicated in red. We then show on the right the result of a large-scale fluid
motion
which brings the two line-vectors together to the same point, stretched and
rotated
by the flow. (It is important to understand that the velocity at scales
$r>\ell_\eta$ cannot
actually bring the two lines to touch, but only within distance $\sim
\ell_\eta$ of each other;
it is then the action of two independent Brownian motions which completes their
transport to the same point.) Around the time of first contact shown in the
figure,
the two line-vectors are positively correlated. Further rotation by the flow
can eventually
produce anti-alignment and negative correlation.  The situation is exactly the
opposite,
however, for the initially longitudinal line-vectors shown in Fig.~12, which
start on the same
(or nearly the same) field-line. As these vectors are brought together by a
similar fluid
motion as before, the line vectors are anti-aligned near the time of first
contact.
They may then become positively oriented as further rotation and stretching
occurs.
These geometric considerations may help to explain why, for large enough $r,$
positive
alignment dominates in $\wh{G}_N$  while negative alignment dominates in
$\wh{G}_L.$
Transverse line-vectors seem to be brought together from large distances by
mainly
lateral advection of field-lines, while longitudinal vectors are brought
together by
twisting and looping of field-lines.

\section{Uniform Fields and Magnetic Induction} \lb{induction}

We now turn to the second question left open in I, the fate of
spatially-uniform
magnetic seed fields in the dynamo regime of the Kazantsev model. This issue
is closely related to the problem of ``magnetic induction'', which is often
defined
simply as the generation of small-scale magnetic fluctuations by turbulent
tangling of the lines of a non-vanishing mean-field. In the case of homogeneous
statistics---which we only consider in this paper---any non-zero average field
must be
spatially uniform. Thus, the response of a uniform field to homogeneous
turbulence
is intimately related to that of magnetic induction. In fact, we shall argue
that there is
no real distinction in this context between ``magnetic induction'' and
``fluctuation dynamo''
with a uniform seed field.

Let us consider then the special case that the initial magnetic-field
$\bB_{(0)}$ is
spatially-uniform (but still possibly random). Applying eq.(\ref{Ben-R-hom})
gives
the energy in the fluctuation field $\bb=\bB-\langle\bB\rangle$ to be
 \begin{eqnarray}
 && \langle b^2(t)\rangle =  \mathcal{R}_{k\ell}(t)\langle
B^k_{(0)}B^\ell_{(0)}\rangle
 -\delta_{k\ell}\langle B^k_{(0)}\rangle\langle B^\ell_{(0)}\rangle \cr
 && =\mathcal{R}_{k\ell}(t)\langle b^k_{(0)}b^\ell_{(0)}\rangle
 + [\mathcal{R}_{k\ell}(t)-\delta_{k\ell}]\langle B^k_{(0)}\rangle\langle
B^\ell_{(0)}\rangle.
 \,\,\,\,\,\,\,\,
 \lb{ben-R-hom} \end{eqnarray}
 In the second line of the equation above, one may make a formal distinction
between
 ``fluctuation dynamo'' represented by the first term and ``magnetic
induction'' represented
 by the second term. However, their asymptotic growth rates at long times are
the
 same, both  given by
 $$ \mathcal{R}_{k\ell}(t)\sim e^{-\lambda t} \mathcal{\wh{G}}_{k\ell},
       \,\,\,\,\,\,\,\,\,\,\,\, (\lambda<0) $$
if the integral
$$  \mathcal{\wh{G}}_{k\ell}\equiv \int d^3r\,\,\mathcal{\wh{G}}_{k\ell}(\br)
$$
is non-vanishing.  We show now that it is indeed non-zero.

Since the above integral is gauge-invariant, one may just as well consider the
integral of $\mathcal{G}_{k\ell}(\br).$ Furthermore, as long as the turbulent
velocity
field (but not necessarily the magnetic field) is statistically isotropic, then
$$ \mathcal{G}_{k\ell} =  \int d^3r\,\,\mathcal{G}_{k\ell}(\br)
      = \frac{1}{3}G_T \delta_{k\ell} $$
where
$ G_T\equiv 4\pi \int_0^\infty dr\,r^2 G_T(r) $
and $G_T(r)=G_L(r)+2G_N(r)$ is the trace function. From (\ref{L-eigfun})
$$  G_T(r)={{\sqrt{2}G_1+G_2}\over{r}}
     = {{2aW_2+(a+\sqrt{2}b)r\partial_rW_2}\over{\lambda r^2}}.$$
Using the incompressibility condition
$a=\sqrt{2}\left(b+{{1}\over{2}}rb'\right)$  and
integrating by parts once in $r$ gives
$$ \int_0^\infty dr \,r^2 G_T(r)=-{{1}\over \sqrt{2}\lambda}
     \int_0^\infty (r^2b''+4rb') W_2(r)\,dr. $$
On the other hand, multiplying the eigenvalue equation (\ref{SL-eq}) by $r^2$
and integrating
by parts in $r$ twice gives
$$ -\int_0^\infty (r^2b''+4rb') W_2(r)\,dr =\lambda \int_0^\infty r^2
W_2(r)\,dr. $$
We finally obtain that
$$ G_T\equiv 4\pi\int_0^\infty dr \,r^2 G_T(r)=2\pi\sqrt{2}
     \int_0^\infty r^2 W_2(r)\,dr>0. $$
We used here the positivity of the ground-state (dynamo) eigenfunction $W_2$.
Notice
that we have not assumed any specific form of $b(r),$ as long as the dynamo
mode exists.

In fact, the entire spatial structure of  small-scale fluctuations for
``magnetic induction'' is
the same as that for ``fluctuation dynamo''  at long times. Both are determined
by the
same dynamo eigenmode. To see this, we note that the magnetic correlation can
be
expanded, in general, into right eigenmodes of $\mathcal{M}:$
 \be  \mathcal{C}^{ij}(\br,t)=\sum_\alpha  c_{\alpha} e^{-\lambda_\alpha t}
      \mathcal{C}^{ij}_\alpha(\br) \lb{C-expand} \ee
The expansion coefficients are determined from the initial correlation
$\mathcal{C}^{ij}_{(0)}(\br)$ by
\be c_{\alpha} \equiv \int d^3r\,\, \mathcal{G}_{\alpha;
ij}(\br)\mathcal{C}^{ij}_{(0)}(\br),
\lb{coeff} \ee
using biorthogonality of right and left eigenmodes $\mathcal{C}_\alpha$ and
$\mathcal{G}_\alpha$.
If $\mathcal{C}^{ij}_{(0)}(\br)=\langle B^i_{(0)}B^j_{(0)}\rangle$ is
$\br$-independent,
then for the ground state mode (say $\alpha=0$)
$ c_0 = \mathcal{G}_{0; k\ell}\langle B^k_{(0)}B^\ell_{(0)}\rangle
     = \frac{1}{3}G_T \langle B_{(0)}^2\rangle>0.$
Thus, with $r$ fixed,
$$   \mathcal{C}^{ij}(\br,t) \sim c_0 e^{-\lambda_0 t} \mathcal{C}^{ij}_0(\br),
  \,\,\,\,\,\,\,\, t\rightarrow \infty.  $$
The magnetic correlation $\mathcal{C}^{ij}(\br,t) $ is dominated by the leading
dynamo eigenmode
$\mathcal{C}^{ij}_0(\br)$  with subleading contributions from additional
eigenmodes (point spectrum of $\mathcal{M}$) if any. There is no essential
physical difference
between the two cases of ``fluctuation dynamo''  ($\bB_{(0)}=\bb_{(0)}$) and
``magnetic induction''
($\bB_{(0)}=\langle\bB_{(0)}\rangle$).

 It is interesting to consider in (\ref{C-expand}) also the opposite limit of
large $r$ with $t$ fixed. In that case, the result with initial condition
$\mathcal{C}^{ij}_{(0)}(\br)
=\langle B^i_{(0)}B^j_{(0)}\rangle$ is
\be  \mathcal{C}^{ij}(\br,t) \sim \langle B^i_{(0)}B^j_{(0)}\rangle,
\,\,\,\,\,\,\,\,r\rightarrow\infty. \lb{cluster} \ee
This is clearly true for the case $\mathcal{C}^{ij}_{(0)}(\br)=\langle
B^i_{(0)}\rangle
\langle B^j_{(0)}\rangle,$ because the above limit then corresponds to
``clustering'' of the
magnetic correlation function. However, the expansion coefficients
(\ref{coeff}) are linear in the
 initial correlation, so the above result must hold whenever
$\mathcal{C}_{(0)}^{ij}(\br)$ is
 $\br$-independent. The limit (\ref{cluster}) comes entirely from the
continuous spectrum
 of $\mathcal{M}$ in the expansion (\ref{C-expand}) (for which the sum over
$\alpha$ is
 actually an integral),  because the finite number of  dynamo eigenmodes
$\mathcal{C}_\alpha$
 associated to the point spectrum of $\mathcal{M}$ all have the
stretched-exponential decay in
 (\ref{W2-large}), with the corresponding value of $\lambda_\alpha<0$. Their
contribution
 thus vanishes in the large-$r$ limit.

 We conclude that there is no essential distinction in the dynamo regime
between
 ``magnetic induction'' for a uniform mean-field and ``fluctuation dynamo'' for
a
 homogeneous (but random) seed field, since all of their physical behaviors and
 underlying mechanisms are exactly the same.  Another possible definition of
 ``magnetic induction'' which is employed in the literature is  the  growth of
small-scale
 fluctuation fields generated from a non-vanishing mean-field, but only in a
parameter
 regime where the ``fluctuation dynamo'' does not exist. The term is used this
way
 in discussion of liquid-metal laboratory experiments operated at a
sub-threshold
 regime \cite{Peffleyetal00,Bourgoinetal02,Nornbergetal06}. Although such
inductive
 growth of fluctuations is generally sub-exponential, it may be exponential if
the
 average magnetic field is itself exponentially growing because of a mean-field
 dynamo effect. Of course, with this definition, it makes no sense to regard
``magnetic
 induction''  and  ``fluctuation dynamo'' as possible co-existing and competing
mechanisms
 for small-scale magnetic field growth.

We may consider as an example of this  second type of ``magnetic induction''
the
failed-dynamo regime of the KK model for $\xi<1,$ at zero Prandtl number and
infinite magnetic Reynolds number. Unlike the liquid-metal experiments where
fluctuation-dynamo fails because $Re_m<Re_m^{(c)},$ the failure here is due to
the extreme roughness of the advecting velocity field (see I). This model
problem
is not a perfect analogue of the induction phenenoma seen in the experiments.
For example, we solve the KK model with velocity integral scale $L=\infty$ (and
thus $Re_m=\infty$) so that the large-scales of our problem are quite different
than
the non-universal, inhomogeneous and anisotropic conditions seen in
experiments.
Also, the average field is uniform in the KK model with homogeneous statistics.
Since spatially constant fields must be time-independent, we cannot study in
this
setting induction of an exponentially growing mean-field. On the other hand,
the cited
experiments  \cite{Peffleyetal00,Bourgoinetal02,Nornbergetal06} have studied
the
turbulent induction of a near-uniform, stationary external field. The analogy
with those
experiments is close enough that we can test some proposed theories of
``magnetic
induction'' in our  model situation.

We therefore consider in this light several results of I for the failed dynamo
regime
of the KK model with $\xi<1.$ It was found there for the case of
spatially-uniform
and isotropic initial data $\mathcal{C}^{ij}_{(0)}(\br)=A\delta^{ij}$ that, at
long times,
\be \langle B^2(t)\rangle \propto \ell_\eta^{\zeta_1}(D_1t)^{|\zeta_1|/\gamma},
\lb{inducedE} \ee
with
$$ \zeta_1=-\frac{3}{2}-\frac{\xi}{2}+\frac{3}{2}
    \sqrt{1-\frac{1}{3}\xi(\xi+2)} $$
which is negative and decreasing from 0 to $-2$ for $0<\xi<1.$ Thus, the energy
in
the magnetic fluctuations is growing, but only as a power-law in $t$. These
same
results hold , in fact, for general uniform initial data of the form
$\mathcal{C}^{ij}_{(0)}(\br)
=\langle B^i_{(0)} B^j_{(0)}\rangle.$ This follows directly from
eq.(\ref{Ben-R-hom})
when the random velocity (but not necessarily the magnetic field) is
statistically
isotropic and the line-correlation matrix
$\mathcal{R}_{k\ell}(t)=\frac{1}{3}\mathcal{R}(t)\delta_{k\ell}.$

Paper I showed further that there are three spatial regimes of the magnetic
correlation.  These are the resistive range:
\be C_L(r,t) \simeq
 A'' \left(\frac{\ell_\eta}{L(t)}\right)^{\zeta_1}\left[ 1-2
\left(\frac{r}{\ell_\eta}\right)^\xi\right]
\,\,\,\,r\ll \ell_\eta, \lb{resist} \ee
the quasi-steady, inertial-convective range:
\be C_L(r,t) \simeq A' \left(\frac{r}{L(t)}\right)^{\zeta_1}\,\,\,\,
\ell_\eta \ll r\ll L(t), \lb{quasi} \ee
and the very large-scale range:
\be C_L(r,t) \simeq A \left[1- \frac{2\zeta_1(\zeta_1+3+\xi)}{\gamma}
           \left(\frac{L(t)}{r}\right)^{\gamma}\right]\,\,\,\, r\gg L(t).
\lb{veryL} \ee
We have here defined $L(t)\equiv (D_1t)^{1/\gamma}$ to be a characteristic
large
length-scale of the magnetic field. Note that $A',A''$ are constants
numerically
proportional to $A.$ Result (\ref{resist}) follows from the formula for
$\Gamma_{in}(\sigma)$
on p.26 of I, together with the series expansion of the hypergeometric function
${\,\!}_2F_1$
(eq.(75) in I).  Equations (\ref{quasi}) and (\ref{veryL}) follow from I,
eq.(67)
for $\alpha=0$ with $\rho\ll 1$ and $\rho\gg 1,$ respectively.  For the last
case,
the large-argument asymptotics of the Kummer function in \cite{Erdelyi53},
eq. 6.13.1(2), is employed to derive the $1/r^\gamma$ term.

These three ranges all have simple physical descriptions, which are easiest
to discuss based on the corresponding magnetic energy spectra obtained by
Fourier transform. In the resistive range the result is
\be E(k,t) \propto  \frac{\langle B^2(t)\rangle}{\eta} k^{-(1+\xi)}
\,\,\,\,\,\,\,\,\,k\gg k_\eta.
\lb{resist-spectrum} \ee
(Note that $\eta=D_1\ell_\eta^\xi.$) This is a Golitsyn-like spectrum, with
physics exactly
like that discussed earlier for the resistive range of the dynamo growth modes.
This range is very universal in the small-$Pr_m$ limit.

The inertial-convective range has energy spectrum
\be E(k,t) \simeq A' [L(t)]^{|\zeta_1|}k^{|\zeta_1|-1}
\,\,\,\,\,\,\,\,\,k_L(t)\ll k\ll k_\eta
\lb{IC-spectrum} \ee
with $k_L(t)=2\pi/L(t)\rightarrow 0$ as $t\rightarrow\infty.$ This growing
range is
responsible for the energy increase in (\ref{inducedE}). It is ``quasi-steady''
in the sense that all its statistical characteristics are identical to those in
the forced
steady-state (see \cite{Vergassola96} and I), with the uniform initial field
replacing
the role of the external force. The physics involves a nontrivial competition
of
stretching $\bB\bdot\grad\bu$ and nonlinear cascade $\bu\bdot\grad\bb.$ If one
assumes that ``induction'' of the initial, background field
$\bB_{(0)}\bdot\grad\bu$
dominates the stretching, then this is  the same physics invoked by
Ruzma\u{\i}kin and
Shukurov \cite{RuzmaikinShukurov82} for induced fluctuations. Assuming a
balance of the terms  $(\bu\bdot\grad)\bb\simeq \bb\bdot\grad\bu\simeq
\bB_{(0)}\bdot\grad\bu,$
they proposed on dimensional grounds an $A k^{-1}$ spectrum, with $A\propto
\langle
B^2_{(0)}\rangle.$ However, such a spectrum only occurs in the KK model for
$\xi=0.$
In that case, the cascade term $\bu\bdot\grad\bb$ can be represented as an
eddy-diffusivity $2D_1$ and a balance equation
$$ \partial_t \bb -2D_1\triangle\bb =\bB_{(0)}\bdot\grad\bu $$
predicts the correct spectrum $\langle B_{(0)}^2\rangle k^{-1}.$ There is no
distinction
in this case between the inertial-convective range and the resistive range with
Golitsyn spectrum.
For $0<\xi<1,$ however,  there is an ``anomalous dimension''  $|\zeta_1|$ which
corrects
the dimensional prediction. There is no longer any simple, heuristic argument
for the magnetic
spectral exponent in the quasi-steady range, which requires the computation of
a
non-perturbative zero-mode. The physics involves a nontrivial competition
between
creation of magnetic energy by stretching and destruction by turbulent cascade
to the
resistive range.

For the very low wavenumber range $k\ll k_L(t)$ one might expect the
dimensional
argument of Ruzma\u{\i}kin and Shukurov to work after all and to correctly
yield an
$Ak^{-1}$ spectrum. The results in (\ref{resist})-(\ref{veryL}) correspond to a
member of the one-parameter family of self-similar decay solutions found in I,
with
parameter choice $\alpha=0.$ For the general member of this family, the
low-wavenumber
spectrum is of the form $A k^{\alpha-1}$ (``permanence of large eddies''; see
I), so setting
$\alpha=0$ should naively lead to the Ruzma\u{\i}kin-Shukurov prediction. But
this is not the case.
Fourier transformation of the initial uniform magnetic field instead yields a
term in the energy
spectrum $\propto A \delta(k),$ a delta-function at $k=0.$ The actual energy
spectrum
in the low-wavenumber range is found by Fourier transforming the second term
in (\ref{veryL}) to be
\be E(k,t) \simeq A' [L(t)]^\gamma k^{1-\xi},  \,\,\,\,\,\,\,\,\, k\ll k_L(t).
\lb{VL-spectrum} \ee
The simple physics of this range is direct induction from the
spatially-constant field
via the balance
$$ \partial_t \bb = \bB_{(0)}\bdot\grad\bu. $$
Indeed, solving this linear Langevin equation yields $E(k,t)\propto \langle
B^2_{(0)}\rangle
(D_1 t) k^2 E_u(k),$ reproducing the above spectrum. Note that
$[L(t)]^\gamma=D_1t$
corresponds to diffusive spectral growth in this range, due to the white-noise
in time character
of the advecting velocity field. The total energy in this range remains
constant in time,
however, because shrinking of the range exactly compensates for increase of the
energy
spectrum. Once the energy $b_k^2\simeq k E(k,t)$ in an interval around
wavenumber $k$ approaches
the energy $\langle B_{(0)}^2\rangle$ in the initial uniform field a nonlinear
cascade
begins and that wavenumber $k$ joins the inertial-convective range with
spectrum
(\ref{IC-spectrum}).

The above argument resembles one which has been invoked to explain a $k^{-5/3}$
spectrum of magnetic energy reported at wavenumbers $k<k_\eta$ in some
low-$Pr_m$
liquid-metal experiments \cite{Peffleyetal00,Nornbergetal06} and in a related
numerical
simulation \cite{Baylissetal07}. These works study the induction of an imposed
magnetic field by a turbulent flow with a Kolmogorov energy spectrum.  The
argument
made by those authors amounts to assuming a balance between induction of
the external magnetic field $\bB_{(0)}$ and convection by the large-scale
velocity
$\bu_{(0)}$ (both mean and fluctuating components):
$$ \bu_{(0)}\bdot\grad\bb = \bB_{(0)}\bdot\grad\bu. $$
This leads to the prediction that $\langle u_{(0)}^2\rangle E(k)\simeq
\langle B_{(0)}^2\rangle E_u(k),$ so that $E_u(k)\simeq
\varepsilon^{2/3}k^{-5/3}$
implies a similar magnetic energy spectrum $E(k)$. We find this argument
unconvincing.
In the first place, large-scale advection conserves the energy in magnetic
fluctuations.
It cannot therefore balance the input from induction  of $\bB_{(0)}.$ If
nonlinear
cascade can be ignored, then magnetic energy must be expected to grow, as in
our
result (\ref{VL-spectrum}) above, and not to saturate. Furthermore, if the
large-scale
sweeping were important, then it should appear in our balance argument for
(\ref{VL-spectrum}), because the KK model contains such dynamical sweeping
effects.
However, we see that the correct spectrum at very low wavenumbers is obtained
by ignoring such sweeping as irrelevant. This is consistent with the results of
Frisch and Wirth \cite{FrischWirth96} for the diffusive range of a passive
scalar,
who emphasized the (nontrivial) fact that sweeping by large-scales plays no
role
in the balance for that high-wavenumber range. Finally, we note that $k^{-5/3}$
is the (Obukhov-Corrsin) spectrum expected for a passive {\it scalar} in a
Kolmogorov
inertial-range and not for a passive magnetic field. The argument of
\cite{Peffleyetal00,
Nornbergetal06,Baylissetal07} thus omits the effects of the small-scale
stretching
interaction $({\bf b}\bdot\grad)\bu$ and we know of no good justification for
doing so
in a steady-state, saturated regime.

\section{Conclusions} \lb{conclusions}

The main purpose of this paper was to quantify the importance of stochasticity
of flux-line freezing to the zero-Prandtl-number turbulent dynamo. It is a
rigorous
result for the  Kazantsev-Kraichnan model that Lagrangian trajectories becomes
intrinsically stochastic in the inertial range with velocity roughness exponent
$\xi$,
due to Richardson 2-particle turbulent diffusion. In this situation,
infinitely-many magnetic
lines in the initial seed field at time $t_0$ are brought to a point at time
$t$
from a region of size $L(t)\sim (t-t_0)^{1/(2-\xi)}.$ Note that the size of the
region
sampled is independent of the resistivity. Not all of the magnetic field lines
in this
large region will, however, make an equal contribution to the net dynamo growth
of magnetic energy. The contribution from lines initially separated by distance
$r$
is quantified, in isotropic non-helical turbulence, by the line-correlations
$R_L(r,t)$
and $R_N(r,t).$ At long times, these correlations are proportional to
$e^{-\lambda t}
\wh{G}_L(r)$ and $e^{-t\lambda t}\wh{G}_N(r),$ respectively, where
$\wh{G}_L(r)$
and $\wh{G}_N(r)$ are longitudinal and transverse components of the
(left/adjoint)
dynamo eigenfunction. The central results of this paper are these two
functions, plotted
in Fig.~10, which quantify the relative contribution to magnetic energy from
lines initially
separated by the distance $r,$ in resistive units.

The principal contribution to the dynamo growth comes from lines at separations
of the order of magnitude of the resistive length $\ell_\eta.$ However, the
decay
of the correlations, in units of the resistive length, is a slow, stretched
exponential.
This implies that lines which arrive from any arbitrary, fixed separation
$r$---no
matter how large---will contribute an amount of energy growing exponentially
rapidly in time. Of course lines separated by $r$ much, much greater than
$\ell_\eta$
will make a small relative contribution, but even lines separated by many
$\ell_\eta$
make a substantial contribution. To underline this fact, we plot in Fig.~13
below
the eigenfunctions $\wh{G}_L(r)$ and $\wh{G}_N(r)$ multiplied by $r^2,$ which,
integrated over $r,$ give the mean magnetic energy for a uniform seed field.
(In order
to make this plot, we have extended our numerical results to $r>163\ell_\eta$
with the
asymptotic  formulas (\ref{G-large-II}).) As this figure should make clear,
lines separated
by many hundreds of resistive lengths are important to the dynamo growth. To be
more quantitative, we find that lines initially separated by up to
$968\ell_\eta$ must
be considered in order to get 90\% of the total magnetic energy. Another
feature
dramatically illustrated in Fig.~13 is the strong {\it anti-dynamo} effect
arising from
lines-vectors initially parallel to the separation vector $\br$, represented by
the
long negative tail in $r^2\wh{G}_L(r).$ We have proposed a heuristic
explanation
for this interesting effect in terms of twisting and looping of field lines
(Figs.~11 and 12).

\begin{figure}[h]
\centerline{\includegraphics[width=9cm,height=7cm]{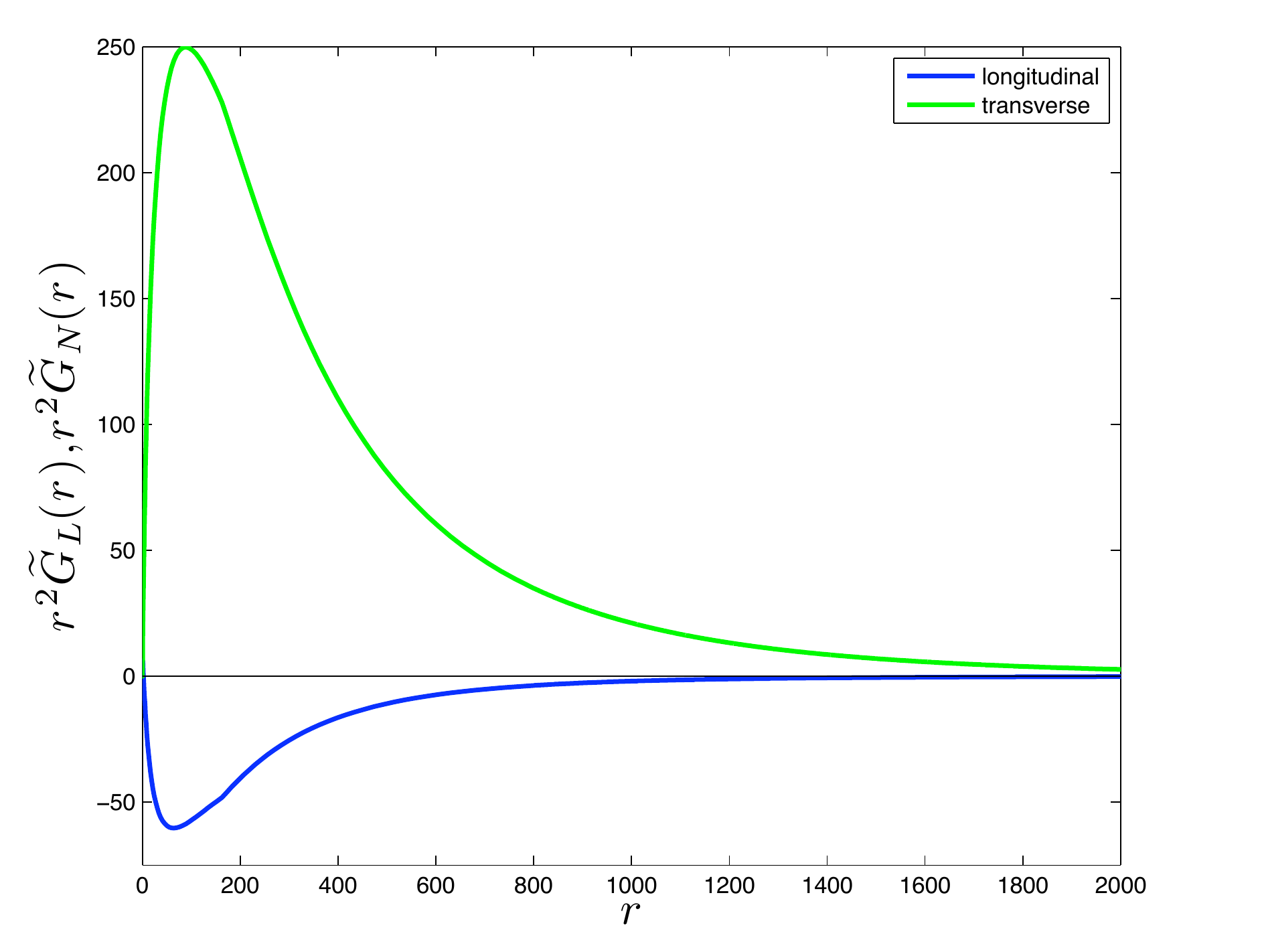}}
\label{rsq_tildeG-fig}
\caption{Contributions of line-correlations to the energy. The integrands,
$r^2\,G_L(r)$ in \textcolor{blue}{blue} and $r^2\,G_N(r)$ in
\textcolor{green}{green},
for magnetic energy with a uniform seed field in
eqs.(\ref{Ben-R-hom}),(\ref{Rint-def}).}
\end{figure}

As we shall see in a following paper \cite{Eyink11} these principle conclusions
apply not only in the Kazantsev-Kraichnan model with $Pr_m=0$ but also to
kinematic dynamo in real hydrodynamic turbulence with $Pr_m\sim 1.$

A second purpose of this paper was to discuss the meaningful distinction,
if any,  between ``fluctuation dynamo'' and ``magnetic induction.'' In the
case of the KK model at $Pr_m=0$ and $Re_m=\infty,$ we have shown that
a uniform mean field (or a uniform random field) may provide the seed field
for small-scale fluctuation dynamo. The asymptotic exponential growth rate
and the small-scale magnetic correlations are exactly the same as for any
other random seed field whose correlation function is non-orthogonal to
the leading dynamo mode. We therefore do not agree with authors
\cite{BrandenburgSubramanian05,Schekochihinetal07,SchuesslerVoegler08}
who distinguish between ``fluctuation dynamo'' and ``magnetic induction''
as two fundamentally different mechanisms.

For example, Sch\"ussler and
V\"ogler \cite{SchuesslerVoegler08} have proposed to explain small-scale
magnetic fields in the quiet solar photosphere as a consequence of near-surface
dynamo action. They mention magnetic induction as a possible alternative
explanation, but dismiss it with the remark: `` `Shredding'  of pre-existing
magnetic flux
(remnants of bipolar magnetic regions) cannot explain the large amount of
observed
horizontal flux since the turbulent cascade does not lead to an accumulation of
energy (and generation of a spectral maximum) at small scales. On the other
hand,
such a behavior is typical for turbulent dynamo action.'' We find in the KK
model that,
quite to the contrary,  induction of a mean field produces the same small-scale
fluctuations and energy spectra as does the fluctuation dynamo. We may indeed
say that---at least in the kinematic regime of weak fields--- ``magnetic
induction''
is nothing but ``fluctuation dynamo'' with a large-scale, deterministic  (mean)
seed
field.

Although Schekochihin et al. \cite{Schekochihinetal07} do make a distinction
between two separate mechanisms, their final conclusion is not so different
from ours.
In their Fig.~8(a) they find, in a sub-threshold regime with $Re_m<Re_m^{(c)},$
that
saturated energy spectra for induction from a uniform field and for decaying
spectra
of the ``failed'' dynamo state {\it exactly} coincide, when normalized to the
same
total energy. They conclude that ``the same mechanism is responsible for
setting the
shape of the spectrum of the magnetic fluctuations induced by a mean field and
of the
decaying or growing such fluctuations in the absence of a mean field,'' just as
we
see in the KK model.

We have also studied the KK model with velocity roughness exponent $\xi<1,$
where the $Pr_m=0$ fluctuation dynamo does not exist, and with a uniform
magnetic
seed field in order to get some insight into the physics of the induction
mechanism in
a failed-dyamo regime. This is very roughly the same situation that was studied
in several
liquid-sodium experiments  with $Pr_m\lesssim 10^{-5}$
\cite{Odieretal98,Bourgoinetal02,
Peffleyetal00,Spenceetal06,Nornbergetal06} and in related numerical simulations
\cite{Baylissetal07,Schekochihinetal07} at somewhat larger Prandtl numbers.
Some of those studies have reported observing a $k^{-5/3}$ spectrum of magnetic
fluctuations in the velocity inertial-range \cite{Peffleyetal00,Nornbergetal06,
Baylissetal07}, while others have reported a $k^{-1}$ spectrum
\cite{Bourgoinetal02,
Schekochihinetal07}. The very least we can conclude from our analysis is that
the explanations offered for those spectra based on inertial-range turbulence
physics do not successfully explain the results in the KK model, even where
those
explanations should apparently apply. An exception is the Golitsyn-Moffatt
argument
for the spectrum  at wavenumbers $k>k_\eta,$ which succeeds (with slight
modification)
in the KK model. At intermediate wavenumbers $k_L(t)<k<k_\eta,$ we find in the
KK model
a range with input of magnetic energy by induction of the mean field balanced
by nonlinear
stretching and cascade to the resistive scale. This is the same physics invoked
in the theory
of Ruzma\u{\i}kin and Shukurov \cite{RuzmaikinShukurov82}, who predicted a
spectrum
$\langle B^2_0\rangle k^{-1}$ on dimensional grounds.  However, this prediction
is
only verified in the KK model for $\xi=0,$ where nonlinear cascade can be
accurately
modelled as an eddy-diffusivity. For the KK model with $0<\xi<1$ this
prediction is
modified by a large anomalous exponent $|\zeta_1|$ (with $|\zeta_1|=2$ for
$\xi=1$)
which must be calculated by a non-perturbative argument. At low wavenumbers
$k<k_L(t)$
in the KK model we find a range with continuous growth of magnetic energy
spectrum
supplied by induction of the uniform field. Unlike the argument in
\cite{Peffleyetal00,
Nornbergetal06,Baylissetal07}, there is no balance with large-scale advection,
although such effects exist in the model. Indeed, it is very unclear to us how
large-scale advection, which conserves magnetic energy, can by itself balance
the
input from induction.

We shall not attempt here to explain in detail the induced spectra observed in
the
experiments and simulations, except to note that they must involve large-scale
physics
outside inertial-range scales. This was, in fact, the point of view of Bourgoin
et al.
\cite{Bourgoinetal02} who explained their $k^{-1}$ spectrum by global
fluctuations
of the flow pattern. Large-scale effects are certainly necessary to obtain
saturation
of the energy. If the length-scale $L(t)$ in (\ref{IC-spectrum}) had a finite
limit for
large times, e.g. for turbulence confined to a box of size $L_B,$ then the
system
should reach a statistical steady-state, just as in the randomly-forced case
\cite{Vergassola96}.
Otherwise, magnetic energy would continue to grow without bound within a
kinematic
description. In such a steady-state, the only extended spectral ranges would be
the nonlinear
range (\ref{IC-spectrum}) and the Golitsyn range (\ref{resist-spectrum}).  The
latter is
well-confirmed in experiments \cite{Odieretal98,Bourgoinetal02,Nornbergetal06}
and
simulations \cite{Baylissetal07,Schekochihinetal07}. If the $k^{-1}$ or
$k^{-5/3}$ spectra
observed at larger scales can be understood at all in terms of inertial-range
turbulence,
then it seems that an analogue of the nonlinear range (\ref{IC-spectrum}) is
the most likely
explanation. If so, then the determination of the precise exponent is a hard
theoretical problem,
for there may be large anomalous scaling corrections to the dimensional
Ruzma\u{\i}kin-Shukurov  $k^{-1}$ spectrum.

The precise spectral exponent of induced magnetic fluctuations in the liquid
metal experiments is open to debate. Over the limited scaling ranges available
it is quite difficult to distinguish between $-5/3$ and $-1$ exponents and the
empirical
spectra can be fit about equally well with both power-laws, or others as well.
It is not even
clear that the concept of  ``scaling exponent'' is entirely well-defined
without the possibility
to extend the putative scaling ranges. Our three spectral ranges
(\ref{resist-spectrum}),
(\ref{IC-spectrum}),(\ref{VL-spectrum}) can be made arbitrarily long by
adjusting parameters,
but, in the experiments and simulations, only the Golitsyn spectral range can
be lengthened
by lowering $Pr_m.$  The low-wavenumber spectral ranges can be made longer only
by increasing $Re_m,$ but, beyond a critical value $Re_{m}^{(c)},$ the dynamo
effect
sets in and the physical phenomena change.

\newpage

\appendix

\section{Boldyrev-Cattaneo-Rosner formalism}\lb{BCR}

We consider in this appendix kinematical relations between the two-point
correlations
of magnetic fields and vector potentials, first for spatially homogeneous
ensembles
and then for both homogeneous and isotropic ensembles (but not
reflection-symmetric).
The results are closely related to those in classical references on homogeneous
turbulence,
such as \cite{Batchelor53} and \cite{MoninYaglom75}, but we employ a
mathematical
reformulation of \cite{Boldyrevetal05} (hereafter, BCR) that was developed for
helical
turbulence. We discuss here, in particular, the consequences of
gauge-invariance.

The two-point correlation of the magnetic field in a homogeneous ensemble
is defined as $\mathcal{C}^{ij}(\br,t)=\langle B^i(\br,t)B^j(\bzed,t)\rangle.$
This satisfies the solenoidality conditions
$\partial_i\mathcal{C}^{ij}=\partial_j
\mathcal{C}^{ij}=0.$ The two-point correlation of the magnetic vector-potential
is likewise $\mathcal{G}_{ij}(\br,t)=\langle A_i(\br,t)A_j(\bzed,t)\rangle$. A
pure gauge field
$A_i^{g}=\partial_i\lambda$ has correlation
$\mathcal{G}_{ij}^g=-\partial_i\partial_j\Lambda$
with $\Lambda(\br)=\langle\lambda(\br)\lambda(\bzed)\rangle.$ Solenoidal
correlations
$\mathcal{C}$ and pure-gauge correlations $\mathcal{G}^g$ form mutually
orthogonal
subspaces in the Hilbert space with inner product
\be \langle \mathcal{C},\mathcal{G}\rangle= \int d^dr\,
\mathcal{C}^{ij}(\br)\mathcal{G}_{ij}
(\br). \lb{in-prod} \ee
This observation will prove important in what follows. Notice also that the
magnetic
and vector-potential fields are related by curls, so that
$$ \mathcal{C}^{ij} = -\mathcal{D}^{ij,k\ell}\mathcal{G}_{k\ell}$$
with the non-positive, self-adjoint differential operator
$$ \mathcal{D}^{ij,k\ell} = \epsilon^{ikp}\epsilon^{j\ell
q}\partial_p\partial_q. $$
It is furthermore useful to introduce the joint correlations of magnetic fields
and
vector-potentials
$$ \Psi^{i\,\,\,\,}_{\,\,\,\,k}(\br,t) = \langle B^i(\br,t)A_k(\bzed,t)\rangle,
\,\,\Psi_{k\,\,\,\,}^{\,\,\,\,i}(\br,t) = \langle
A_k(\br,t)B^i(\bzed,t)\rangle$$
which are related by
$\Psi^{i\,\,\,\,}_{\,\,\,\,k}(\br,t)=\Psi_{k\,\,\,\,}^{\,\,\,\,i}(-\br,t).$
These are obtained from the vector-potential correlation by
$$\Psi^{i\,\,\,\,}_{\,\,\,\,\ell}=\mathcal{R}^{i,k}\mathcal{G}_{k\ell} $$
with
$$ \mathcal{R}^{i,k} = \epsilon^{i p k}\partial_p,$$
and likewise
$$\Psi_{k\,\,\,\,}^{\,\,\,\,j}=-\mathcal{R}^{j,\ell}\mathcal{G}_{k\ell}
           = (\mathcal{R}^{j,\ell})^*\mathcal{G}_{k\ell}. $$
The symbol * here denotes adjoint with respect to $\langle f,g\rangle=\int
d^3r\, f(\br)g(\br).$
Note that $\mathcal{D}^{ij,k\ell} =  - \mathcal{R}^{i,k}
(\mathcal{R}^{j,\ell})^*.$

We are mainly concerned with statistics both homogeneous and isotropic, but not
necessarily invariant under space reflections. In that case, the magnetic
correlation
becomes
$$\mathcal{C}^{ij}(\br,t)=C_L \hat{r}^i\hat{r}^j +
     C_N (\delta^{ij}-\hat{r}^i\hat{r}^j ) + C_H \epsilon^{ijk}\hat{r}_k, $$
with coefficients depending only on $r,t,$ and the solenoidal condition becomes
$$ C_N(r,t) = C_L(r,r) + \frac{1}{2}rC_L'(r,t). $$
Note that there is no condition on $C_H,$ as the third term is always
divergence-free.
One can similarly write the vector-potential correlation as
$$\mathcal{G}_{ij}(\br,t)=G_L \hat{r}_i\hat{r}_j +
     G_N (\delta_{ij}-\hat{r}_i\hat{r}_j ) + G_H \epsilon_{ijk}\hat{r}^k. $$
For a pure-gauge field
$$ G_L^g(r,t)=\Lambda''(r,t),\,\,\, G_N^g(r,t)=\frac{1}{r}\Lambda'(r,t),\,\,\,
     G_H^g(r,t)\equiv 0. $$
This is equivalent to the conditions
$$ G_L(r,t)=G_N(r,t)+rG_N'(r,t), \,\,\,\, G_H(r,t)\equiv 0. $$

We consider also the mixed magnetic and vector potential correlators:
$$\Psi^{i\,\,\,\,}_{\,\,\,\,j}(\br,t)=\Psi_L \hat{r}^i\hat{r}_j +
     \Psi_N (\delta^{i\,\,\,}_{\,\,\,j}-\hat{r}^i\hat{r}_j ) +
     \Psi_H \epsilon^{i\,\,\,}_{\,\,\,jk}\hat{r}^k, $$
This correlation must be divergence-free in both indices $i$ and $j.$
This is trivial for $i$ and follows in $j$ for the first two terms by their
symmetry in $i,j.$ The third antisymmetric term is always solenoidal in
both indices.  Another interesting fact is that this mixed correlation is
gauge-invariant in the isotropic sector. To see this, note that
$$ \langle B^i(\br,t)\Lambda(\bzed,t)\rangle= P(r,t) \hat{r}^i $$
by isotropy. The solenoidal condition implies that $P(r,t)=\alpha(t)/r^2$ and
regularity
for $r\rightarrow 0$ implies that $\alpha(t)\equiv 0.$ By differentiation it
then follows that
also
$$ \langle B^i(\br,t)\partial_k\Lambda(\bzed,t)\rangle= 0, $$
which implies gauge-invariance of $\Psi^{i\,\,\,\,}_{\,\,\,\,k}$. The same
argument implies
also that every vector-potential correlation in the isotropic sector can be
written as a sum
$\mathcal{G}=\mathcal{G}^s+\mathcal{G}^g$ of a solenoidal correlation
$\mathcal{G}^s$
and pure-gauge correlation $\mathcal{G}^g.$ This follows by writing the
vector-potential
field as $\bA=\bA^s+\grad\Lambda,$ where $\bA^s$ is the Coloumb-gauge potential
satisfying $\grad\bdot\bA^s=0,$ and $\mathcal{G}^s$ is thereby identified as
the
correlator of the Coloumb-gauge vector potential field.

The relation
$\Psi^{i\,\,\,\,}_{\,\,\,\,\ell}=\mathcal{R}^{i,k}\mathcal{G}_{k\ell}$ becomes
in the isotropic sector
$$\Psi_L=2 \frac{G_H}{r},\,\,
\Psi_N = \frac{G_H}{r}+G_H',\,\, \Psi_H = \frac{G_L-G_N}{r}-G_N'. $$
The solenoidality condition $\Psi_N =\Psi_L+\frac{1}{2}r\Psi_L'$ is
satisfied automatically, consistent with our earlier conclusion. The relation
$\mathcal{C}^{ij}=\mathcal{R}^{i,k}\Psi_{k\,\,\,\,}^{\,\,\,\,j}$
likewise becomes
$$ C_L=2 \frac{\Psi_H}{r},\,\,
     C_N = \frac{\Psi_H}{r}+\Psi_H',\,\,
     C_H = \frac{\Psi_L-\Psi_N}{r}-\Psi_N' $$
by using
$$\Psi_{k\,\,\,\,}^{\,\,\,\,j}(\br,t)=\Psi_L \hat{r}_k\hat{r}^j +
     \Psi_N (\delta_{k\,\,\,}^{\,\,\,j}-\hat{r}_k\hat{r}^j) +
     \Psi_H \epsilon_{k\,\,\,}^{\,\,\,j\ell}\hat{r}_\ell, $$
with the same coefficient scalar functions as
$\Psi^{i\,\,\,\,}_{\,\,\,\,\ell}.$ Note also that
$C_H=-2\Psi_L'-\frac{1}{2}r\Psi_L''$
by employing the solenoidality relation between $\Psi_N$ and $\Psi_L.$

The main idea of BCR was to employ a slightly different basis,
$$\mathcal{C}^{ij}(\br,t)= C_1 \frac{\delta^{ij}-\hat{r}^i\hat{r}^j
}{\sqrt{2}r}
   + C_2 \frac{\hat{r}^i\hat{r}^j}{r} + C_3
\frac{\epsilon^{ijk}\hat{r}_k}{\sqrt{2}r}, $$
and likewise for $\mathcal{G}_{ij},$ with normalizations chosen so that the
Hilbert space inner product in the isotropic sector becomes
$$ \langle \mathcal{C},\mathcal{G}\rangle
     =4\pi\int_0^\infty dr\, [C_1G_1+C_2G_2+C_3G_3]. $$
Trivially $C_1=\sqrt{2}rC_N,\,C_2=rC_L,\,C_3=\sqrt{2}rC_H.$  The main advantage
of this normalization is the simplification of various differential operators.
A
straightforward calculation (see details below) gives
$\mathcal{C}=-\mathcal{DG}$ with
$$     {\cal D}=\left( \begin{array}{ccc}
        \partial_r^2 & -\partial_r{{\sqrt{2}}\over{r}} &  0\cr
       {{\sqrt{2}}\over{r}}\partial_r & -{{2}\over{r^2}} & 0 \cr
       0 & 0 & {{1}\over{r^2}}\partial_r  r^4\partial_r{{1}\over{r^2}}
       \end{array}\right) $$
acting on the 3-vector function $\mathcal{G}=(G_1,G_2, G_3)^\top.$  The
operator
$\mathcal{D}$ is explicitly self-adjoint in this representation. Similar
simplifications
occur in the operators $\mathcal{R}$ and $\mathcal{R}^*.$ Following BCR we
introduce
$$ \mathcal{W}=\left(\begin{array}{c}
                                       0 \cr
                                       W_2 \cr
                                       W_3
                                       \end{array}\right), $$
with
$$ W_2=\Psi_3=\sqrt{2}r\Psi_H, \,\,\,\,\,
     W_3=\Psi_1-\sqrt{2}\Psi_2=\frac{1}{\sqrt{2}}r^2 \Psi_L'. $$
This representation exploits the fact that $\Psi$ has implicitly only two
degrees of
freedom, due to the solenoidality condition on $\Psi_L$ and $\Psi_N.$  A bit of
calculation yields
$$ \mathcal{C}=\mathcal{RW},\,\,\,\,\,\mathcal{W}=\mathcal{R}^*\mathcal{G} $$
with
$$       \mathcal{R}=\left( \begin{array}{ccc}
                                            0 & \partial_r &  0\cr
                                            0 & {{\sqrt{2}}\over{r}} & 0 \cr
                                            0 & 0 & -{{1}\over{r^2}}\partial_r
r^2 \cr
                                            \end{array}\right), \,\,\,\,\,\,
        \mathcal{R}^*= \left( \begin{array}{ccc}
                                            0 & 0  &  0\cr
                                            -\partial_r & {{\sqrt{2}}\over{r}}
& 0 \cr
                                            0 & 0 &
r^2\partial_r{{1}\over{r^2}} \cr
                                            \end{array}\right), $$
The range of the operator $\mathcal{R}$ is easily checked to consist of the
solenoidal
correlations, while its adjoint $\mathcal{R}^*$ annihilates pure-gauge fields,
associated
to gauge-invariance of $\mathcal{W}.$ Combining the above results gives the
previous
formula for $\mathcal{D}=-\mathcal{RR}^*.$ We see that $\mathcal{D}$ is a
negative-definite,
self-adjoint operator in the isotropic sector.

\section{Numerical Methods}\lb{numerics}

This appendix discusses the numerical solution of the equation (\ref{Lam-eq2})
in the text:
$$ \partial_r(r^2b\partial_r\Lambda)+\lambda r^2\Lambda=- r^2 F. $$
The output of {\tt MATSLISE} provides an input to this equation, via the source
term $F=2\Phi+TG.$ We shall therefore employ the same grid of points $r_i,\,\,
i=1,...,L$ that was generated by the {\tt MATSLISE} algorithm for solving the
Sturm-Liouville problem (\ref{SL-eq-p}). It is useful to observe that that grid
is
approximately evenly spaced on a logarithmic scale. Thus, we introduce
into (\ref{Lam-eq2}) the variable $\sigma=\ln r$ and multiply by $r,$ giving
\be
\partial_\sigma(p(\sigma)\partial_\sigma\Lambda)+q(\sigma)\Lambda=-f(\sigma)
\lb{Lam-eq3} \ee
with definitions (different than in (\ref{SL-eq-p}))
$$ p = e^\sigma b(e^\sigma),\,\,\,\, q=\lambda e^{3\sigma},\,\,\,\,
f=e^{3\sigma}F(e^\sigma). $$
We then discretize (\ref{Lam-eq3}) as
\begin{eqnarray*}
&& \frac{1}{\delta\sigma_i}\left[
p(\sigma_i)\left(\frac{\Lambda_{i+1}-\Lambda_i}{\delta\sigma_i}\right)
-p(\sigma_{i-1})\left(\frac{\Lambda_i-\Lambda_{i-1}}
{\delta\sigma_{i-1}}\right)\right] \cr
&&
\,\,\,\,\,\,\,\,\,\,\,\,\,\,\,\,\,\,\,\,\,\,\,\,\,\,\,\,\,\,\,\,\,\,\,\,\,\,\,\,\,\,\,\,
    +q(\sigma_i)\Lambda_i = -f(\sigma_i)
\end{eqnarray*}
with $\delta\sigma_i=\sigma_{i+1}-\sigma_i$ and $\sigma_i=\ln r_i,\,\,\,
i=1,2,...,L$ .  An extra point is added with $\sigma_{L+1}=\infty$ and
$\Lambda_{L+1}=0$
to define the components with $i$ or $j=L.$ This leads to the $L$-vector
equation
\be {\bf M}\boLambda={\bf F} \lb{Lam-eq-disc} \ee
with negative-definite, self-adjoint, tridiagonal matrix
\begin{eqnarray*}
&& M_{ij}=\frac{p(\sigma_i)}{\delta\sigma_i}\delta_{j,i+1}+
\frac{p(\sigma_{i-1})}{\delta\sigma_{i-1}}\delta_{j,i-1}+ \cr
&& \,\,\,\,\,\,\,\,\,\,\,\,\,\,\,\,\,\,\,\,\,\,\,
\left(q(\sigma_i)\delta\sigma_i-\frac{p(\sigma_i)}{\delta\sigma_i}
-\frac{p(\sigma_{i-1})}{\delta\sigma_{i-1}}\right)\delta_{ij}
\end{eqnarray*}
and vector $F_i=-f(\sigma_i)\delta\sigma_i.$ The $i=L$ equation is
$$ q(\sigma_L)\Lambda_L=f(\sigma_L) $$
with direct solution $\Lambda_L=F(\sigma_L)/\lambda.$ The remaining $L-1$
equations
can be solved with the above value as boundary condition on $\Lambda_L$ or,
more simply,
with the boundary condition $\Lambda_L=0$ (since $F(\sigma)\rightarrow 0$ as
$\sigma\rightarrow \infty).$ We have verified that both choices of boundary
conditions
lead to very quantitatively similar results; we thus present only those with
the condition
$\Lambda_L=0$. To calculate the quantity $\Phi$ that appears in the source-term
$F$ we
 evaluate the integral $\int_0^r d\rho \,\,a(\rho)W_2(\rho)/\rho$ using the
composite
 trapezoidal rule ({\tt cumtrapz} in {\tt MATLAB}).  The first $(L-1)$ linear
equations
 from (\ref{Lam-eq-disc}) are then solved using {\tt mldivide} in {\tt MATLAB},
giving
 $\Lambda_i,\,\,i=1,2,...,L-1.$

 Getting corrections $\delta G_1=\sqrt{2}\Lambda'(r),$ $\delta
G_2(r)=r\Lambda''(r)$
 requires us to approximate also derivatives of $\Lambda(r).$ We employ the
formula
$\Lambda'(r)=\frac{1}{r}\partial_\sigma\Lambda$ discretized using central
differences as
$$\Lambda'_i=\exp\left(-\frac{\sigma_{i+1}+\sigma_{i-1}}{2}\right)
\frac{\Lambda_{i+1}-\Lambda_{i-1}}{\sigma_{i+1}-\sigma_{i-1}}. $$
At the endpoints $i=1,L$ corresponding forward and backward-difference
approximations are employed. To calculate the second-derivative $\Lambda_i''$,
the same formulas are employed with $\Lambda_i$ replaced by $\Lambda_i'.$
The difference approximations yields some spurious oscillations for $r\lesssim
2,$
which are removed by filtering.

The parameters $\Lambda_0,\Lambda_1,$ and $\Lambda_s$ in the small-$r$
asymptotics (\ref{Lam-reg}),(\ref{Lam-sing}) for $\Lambda(r)$ were determined
as follows. We found directly that $\Lambda_0 = -1535.5$ by taking the limit of
small $r$ in the numerical solution. We then calculated $\Lambda_1$ in two
different ways. First, we solved eq.(\ref{F0-eq}), or
$\lambda\Lambda_0+12\Lambda_1
=-F_0$. To get a sufficiently accurate value of
$F_0=2\Phi_0+24\sqrt{2}A/\lambda$
we had to calculate an accurate value of the integral
$$ \Phi_0 =-\int_0^\infty dr \frac{aW_2}{r} \doteq -208.1745$$
To get this value, we extended the numerical solution for $W_2$ from {\tt
MATSLISE}
using the large-$r$ asymptotic formula (\ref{W2-large}) from $r=560$ to
$r=10^4,$
and then used a composite trapezoid rule to approximate the integral. This gave
$F_0\doteq 439.043$ and a resulting value $\Lambda_1 =11.9299.$ On the other
hand, we could also determine $\Lambda_1$ by a least-square linear fit to
$\Lambda(r)-\Lambda_0$ in a log-log plot. Over the range from $r=2.7\times
10^{-6}$
to $r=0.12$ the best fit was $\Lambda(r)-\Lambda_0\doteq \Lambda_1 r^m$ with
$m=1.9999$ and $\Lambda_1=11.9517,$ agreeing quite well with the previous,
independent
determination. Finally, we calculated $\Lambda_s =-6.0232$ from the already
determined numerical constants and eq.(\ref{F1-eq}).  It was verified that the
full
small-$r$ asymptotic expression $\Lambda=\Lambda_{reg}+\Lambda_{sing}$
with these constants gave an accurate fit to the numerical result for $r<0.5.$

\begin{acknowledgments}
This work was partially supported by NSF Grant No. AST-0428325 at Johns Hopkins
University
and was completed during the author's spring 2010 sabbatical at the Center for
Magnetic
Self-Organization in Laboratory and Astrophysical Plasmas at the University of
Wisconsin - Madison. We acknowledge the warm hospitality of the center and of
Ellen Zweibel, Center Director.
\end{acknowledgments}

\bibliography{DynamoPr0_PRE}

\end{document}